\documentclass[twocolumn]{aastex63}
\usepackage{amsmath}     
\usepackage{wasysym}           
\usepackage{graphicx}
\usepackage{amssymb}
\usepackage{epstopdf}
\usepackage{mathrsfs}
\usepackage{anyfontsize}
\usepackage{natbib}
\usepackage{color}
\usepackage{lipsum}
\usepackage{diagbox}

\shorttitle{Explosions from failed supernovae}
\shortauthors{Coughlin}
\begin{document}
\title{The division between weak and strong explosions from failed supernovae}
\author[0000-0003-3765-6401]{Eric R.~Coughlin}
\affiliation{Department of Physics, Syracuse University, Syracuse, NY 13210, USA}

\email{ecoughli@syr.edu}

\begin{abstract}
Some massive stars likely fail to produce core-collapse supernovae, but these failed supernovae (FSNe) can generate an electromagnetic outburst prior to the disappearance of the star, as the mass lost to neutrinos during the stellar core-collapse results in the formation and breakout of a second shock. We show that when the mass lost to neutrinos is sufficiently small, there are two self-similar solutions that describe the propagation of a weak shock into a hydrodynamically expanding envelope that simultaneously yield accretion onto the black hole. The larger-Mach number solution is unstable and yields the minimum Mach number that a shock must have to strengthen into the energy-conserving regime. Above a critical mass loss there are no weak-shock solutions, implying that there are only strong explosions if the neutrino mass loss is above a critical value, and this value is a few percent of the mass of the star (and is physically achievable) for typical parameters.  Our results imply that the fate of the explosion from a FSN -- weak with little to no mass ejection or strong with the expulsion of the majority of the envelope -- is a sensitive function of the stellar properties and the neutrino mass loss. We also show that there is a second type of self-similar solution for the shock that results in the ``settling'' of the gas near the compact object, which may be applicable to non-terminal stellar eruptions and the response of a gaseous disc to gravitational-wave induced mass loss from a binary black hole merger. 
\end{abstract}


\section{Introduction}
Core-collapse supernovae (CCSNe) -- during which a massive-star implodes, reverses the implosion, and generates a luminous explosion that outshines the entire galaxy in which it resides -- demarcate the most fantastic outcomes of high-mass stellar death. While historically these explosions were visible only when they occurred within our Galaxy, in recent years and with the advent of survey science (e.g., the Panoramic Survey Telescope and Rapid Response System, Zwicky Transient Facility, the All-Sky Automated Search for Supernovae, Dark Energy Survey, and -- in the imminent future -- Rubin Observatory; \citealt{chambers16, Bellm2017, Shappee2014, Dark2016, Ivezic19}, respectively), we are now detecting multiple CCSNe on a nightly basis. Entirely new classes of supernovae have also been detected, including superluminous (e.g., \citealt{Gal-Yam2012, Dong2016}), relativistic (e.g., \citealt{Galama1998, Soderberg2010b}), and long-lived and variable/interacting supernovae (e.g., \citealt{Arcavi2017, Sollerman2020}), all of which challenge our explosion models and likely require a sustained engine or a distinct production mechanism. 

While nature indubitably generates CCSNe in copious numbers, both theory and observation are consistent with the notion that not all massive stars that undergo core collapse produce successful explosions, and instead some progenitors continue to implode to the point of black hole formation. In particular, hydrodynamical simulations in one (e.g., \citealt{woosley95, oconnor11, ugliano12, sukhbold16}), two (e.g., \citealt{janka96, fryer00, blondin03, nakamura15, vartanyan23}), and three dimensions (e.g., \citealt{couch14, lentz15, takiwaki16, chan18, vartanyan19, vartanyan22, varma23}) have found that certain types of star, and specifically those that exhibit a high concentration of mass near the neutron star at the time of implosion (i.e., a high ``compactness'' parameter; \citealt{oconnor11, sukhbold14, ertl16}), do not revive the accretion shock that forms from the neutron star bounce (i.e., the original mechanism proposed to generate CCSNe; \citealt{colgate60}). Observations also suggest that the massive-star formation rate exceeds the core-collapse rate \citep{horiuchi11}, there is a dearth of supernovae from high-mass stellar progenitors \citep{smartt15}, and one star has been found that has faded in brightness below detectability without a supernova \citep{adams17}. Overall, the core-collapse failure rate, or the number of {failed supernovae} (FSNe), could be on the order of 20-30\% of the stellar death rate, i.e., as many as 30\% of all massive stars could end their lives in failure \citep{kochanek08}. 

While a natural conclusion is that -- in the absence of substantial rotation (which if present could lead to disc formation and the launching of relativistic jets in a collapsar-like scenario; \citealt{woosley93, macfadyen99, quataert12}) -- FSNe are analogous with disappearing stars, the outcome is more subtle. Specifically, it was recognized by \citet{nadezhin80} that, because the neutron star formation during core collapse coincides with the liberation of up to $\sim 0.5 M_{\odot}$ in mass-energy in the form of neutrinos \citep{burrows88}, the abrupt reduction in the central mass of the star corresponds to the outward motion of the overlying envelope. A secondary, relatively weak shock is formed between the imploding inner regions of the star and the expanding outer envelope, which propagates through to the edge of the star and breaks out. Therefore, even FSNe coincide with a less-luminous (than a typical CCSN) eruption that signifies the death of the star.

A number of investigations have since analyzed the physical properties of the shock, its propagation through the overlying envelope, the bulk energetics of the explosion, and the appearance of the eruption as the shock breaks out of the surface \citep{lovegrove13, piro13, lovegrove17, fernandez18, coughlin18, coughlin18b, ivanov21}. Two conclusions that can be drawn from these works are 1) The shock can unbind a substantial fraction of the stellar hydrogen envelope in some cases, but in others the vast majority of the envelope remains bound and the explosion is extremely weak, and which of these two outcomes is realized depends on the neutrino mass loss and the stellar type (which are not decoupled, as stars with high stellar compactness overcome the TOV limit in a correspondingly shorter time, which limits the neutrino mass loss; \citealt{fernandez18}); and 2) The shock is weak, with a Mach number on the order of a few at most. The latter feature implies that the shock propagation is qualitatively distinct from the strong regime, as (e.g.) the binding energy of the envelope and the finite ambient (i.e., pre-shock) gas pressure are dynamically important. 

Despite these added complexities to the shock propagation, it was shown by \citet{coughlin18b} that there exists a distinct (from the Sedov-Taylor blastwave; \citealt{sedov59, taylor50}) class of self-similar solution to the fluid equations that describes the propagation of a weak shock through a power-law (and hydrostatic) ambient medium. These solutions account for the gravitational field of the mass interior to the shock, the finite ambient pressure, and result not just in the outward motion of the post-shock gas, but also the fallback onto the newly formed black hole through a stagnation and sonic point. These weak shocks with accretion were also shown to give good agreement with the numerical solution obtained by \citet{fernandez18} in the specific case of a yellow supergiant (YSG). 

Additional physical insight into the nature of these self-similar solutions was provided by \citet{coughlin19, ro19}, who demonstrated analytically and numerically that 1) The weak shock solutions described in \citet{coughlin18b} are weakly unstable, such that the deviation of (e.g.)~the shock position from the self-similar solution grow as $\propto t^{\alpha}$ in time, with $\alpha \lesssim 0.2$, and 2) There exists a second self-similar solution to the fluid equations that corresponds to a rarefaction wave with Mach number equal to 1, and this second solution was shown to be stable (i.e., perturbations scale as $\propto t^{\alpha}$ with $\alpha < 0$). From these investigations, the conclusion is that the weak-shock solutions found by \citet{coughlin18b}, and in particular the Mach number of the shock, represent the dividing line between asymptotically strong shocks (which accelerate into the Sedov-Taylor/energy-conserving regime) and weak shocks (which decelerate to the rarefaction wave solution). 

While the sequence of papers \citet{coughlin18b, coughlin19, ro19} shed light on the nature of weak shock propagation in FSNe (and in general), there remains the question of how the Mach number of the shock is related to the mass lost to neutrinos. In particular, the self-similar solutions (both the weak-shock solution and the rarefaction wave) assume that the ambient gas into which the shock propagates is in hydrostatic balance. However, as described above, there is an instantaneous and largely dynamical (i.e., the gas responds almost exclusively through a change in its velocity, but see Section \ref{sec:ambient} below) response of the envelope that causes the gas to move out, and this effect changes the shock jump conditions -- and thus the solution for the post-shock gas -- nontrivially.

There is also a question related to the initial formation of the shock: previous analytical estimates \citep{fernandez18, coughlin18} used the fact that there is a location within the progenitor at which the neutrino diffusion time ($\sim 1$ s; \citealt{burrows88}) is comparable to the dynamical time of the gas. For radii smaller than this distance, the gas responds quasi-hydrostatically, and hence the wave that communicates the mass loss should be launched from near this radius and correspondingly steepen into a shock. However, in a collapsing star the inner regions are in dynamical freefall, with the freefall communicated to the overlying envelope by the propagation of a rarefaction wave. It seems possible that the radius within the original progenitor at which the freefall time equals the diffusion time is {already in dynamical freefall}, meaning that the mass loss occurs effectively instantaneously in the region of the star in which we would expect the shock to be able to form (i.e., at radii larger than that coincident with the rarefaction wave generating the infall). The question then becomes: how does the shock form when the mass loss occurs instantaneously, and how does the shock formation occur in concert with the pre-existing rarefaction wave?

It is our goal here to answer these questions. In Section \ref{sec:basic} we first use a simple model that accounts only for the dynamical (i.e., pressure-less) response of the gas to the instantaneous reduction in the gravitational field of a point mass, both to gain intuition as to the behavior of the fluid and also to show that a caustic -- where Lagrangian fluid shells would cross in the absence of pressure -- forms within the expanding flow that moves outward as $\propto t^{2/3}$, where $t$ is time since the (instantaneous) mass loss. We then show in Section \ref{sec:ambient} that there is an exact solution to the fluid equations that describes the response of gas in a power-law and initially hydrostatic medium to an instantaneous reduction in the mass of the gravitating object (i.e., the body responsible for maintaining hydrostatic balance). These solutions approach zero velocity at asymptotically large radii from the central object, and terminate in a sonic point at a time-dependent radius that scales as $\propto t^{2/3}$. The existence of a sonic radius implies that these expanding solutions must match onto a shock at some (time-dependent) location outside of that radius. 

In Section \ref{sec:shock}, we show that there are {two, weak-shock solutions} that match onto the expanding medium through the shock jump conditions. In the limit that the mass loss goes to zero, one solution tends to the rarefaction wave (i.e., Mach number unity), while the other tends to the unstable solution derived in \citet{coughlin18b}, and hence these represent stable and unstable (respectively) generalizations to the self-similar solutions found in \citet{coughlin18b, coughlin19c}. Moreover, we show that these weak-shock solutions exist {only below a critical neutrino mass-loss}, and above this critical mass loss (which itself is a function of the properties of the ambient medium) there is no weak-shock solution; this implies that sufficiently high (and astrophysically relevant, as the critical mass loss is on the order of a few percent of the mass of the star for typical parameters; see Table \ref{tab:Machmax} below) neutrino mass loss results in the formation of a strong shock, substantial mass ejection, and a correspondingly luminous outburst.

We discuss the implications of our findings in the context of failed supernovae in Section \ref{sec:discussion} before summarizing and concluding in Section \ref{sec:summary}. In Appendix \ref{sec:settling}, we also show that there are two self-similar ``settling'' solutions, which remain causally connected throughout the post-shock region and yield zero velocity and zero mass flux near the origin. These solutions may be relevant to non-terminal stellar eruptions and the response of a circumbinary flow to gravitational-wave induced mass loss following the merger of a black hole binary. 

\section{Basic Considerations and Dynamical model}
\label{sec:basic}
To gain some understanding as to the response of a medium to the gravitational mass loss that follows the neutron star formation, assume that the mass loss occurs instantaneously and that the pressure and density are unchanged from their hydrostatic values. In this case, the response of the fluid is purely dynamic, and the equation of motion in Lagrangian form is (assuming that the pressure gradient continues to balance the original gravitational field)
\begin{equation}
\frac{\partial^2r}{\partial t^2} = \frac{G\delta M}{r^2}, \label{dynamic}
\end{equation}
where $\delta M > 0$ is the mass lost to neutrinos and $r$ is the Lagrangian position of a fluid shell. The solution to this equation can be written as
\begin{equation}
r = r_0f(\eta), \quad \eta \equiv \frac{\sqrt{G\delta M}t}{r_0^{3/2}}, \label{rss}
\end{equation}
where $r_0$ is the initial (at $t = 0$) Lagrangian position and the initial conditions (assuming the fluid is originally static) demand $f(0) = 1$ and $df/d\eta(0) = 0$. Then inserting $r = r_0f(\eta)$ into Equation \eqref{dynamic} gives
\begin{equation}
\frac{d^2f}{d\eta^2} = \frac{1}{f^2}.
\end{equation}
For $\eta \simeq 0$ and $f \simeq 1$, this shows that $df/d\eta \simeq \eta$, and hence that $v \simeq G\delta M t/r_0^2$; this is consistent with the findings of \citet{coughlin18}, who obtained the same leading-order (in $\eta$) solution using a perturbative approach.

A shock will form approximately when fluid elements start to cross at a caustic, which occurs when $\partial r/\partial r_0 = 0$, i.e., when two fluid shells with different $r_0$ have the same position. Differentiating Equation \eqref{rss} with respect to $r_0$, this condition is satisfied when
\begin{equation}
f-\frac{3}{2}\eta\frac{df}{d\eta} = 0 \quad \Leftrightarrow \quad \eta \simeq 1.237 \equiv \eta_{\rm sh},
\end{equation}
where $\eta_{\rm sh} \simeq 1.237$ is the numerically obtained value at which this equation is satisfied. The characteristic of the caustic, which we denote $R_{\rm 0, sh}(t)$, is then given by
\begin{equation}
\begin{split}
&R_{0,\rm sh}(t) = \left(\frac{\sqrt{G\delta M}t}{\eta_{\rm sh}}\right)^{2/3} \simeq 0.868 \left(\sqrt{G\delta M}t\right)^{2/3}, \\
&R_{\rm sh}(t) = R_{0,\rm sh}(t)f(\eta_{\rm sh}) \simeq 1.419 \left(\sqrt{G\delta M}t\right)^{2/3}. \label{Rshoft}
\end{split}
\end{equation}

\begin{figure}[htbp] 
   \centering
   \includegraphics[width=0.47\textwidth]{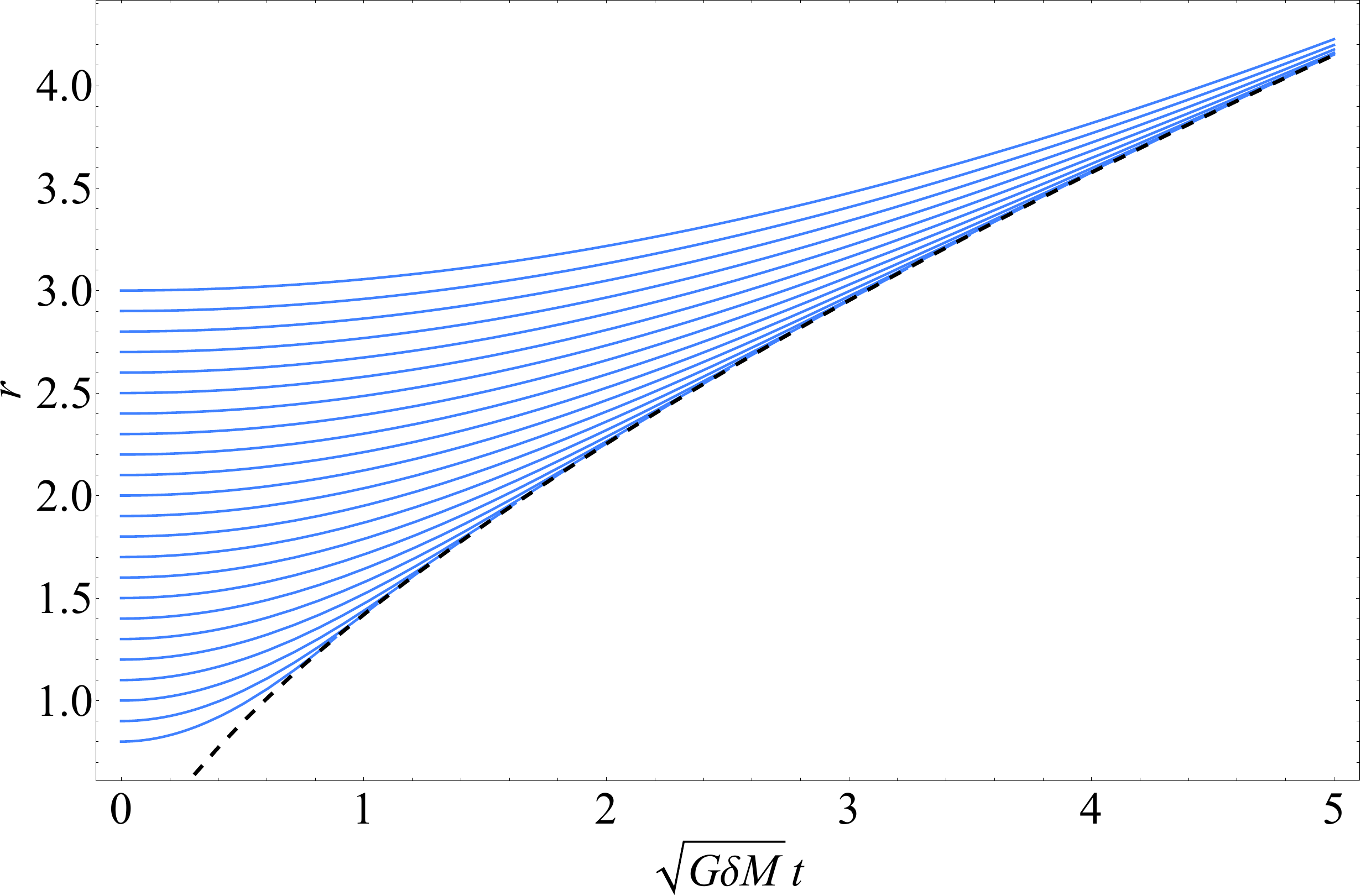} 
   \caption{The characteristics of fluid elements as functions of $\sqrt{G\delta M}t$. The black, dashed curve shows the caustic that forms as fluid shells start to cross, the characteristic of which is given by Equation \eqref{Rshoft}. }
   \label{fig:rlag}
\end{figure}

Figure \ref{fig:rlag} shows the characteristic curves of a number of fluid elements as functions of time following the mass loss. The black, dashed curve illustrates the location of the caustic, given by Equation \eqref{Rshoft}, and coincides with the time-dependent location at which fluid elements first cross. In the absence of pressure, this would coincide with where a shock forms within the flow. If we imagine that there is some inner edge to the envelope, which corresponds to an initial radius $r_{0, \rm i}$, and the initial density profile of the envelope $\rho_{0}(r_0)$ can be approximated as a power-law, such that $\rho_0(r_0) = \rho_{\rm 0, i}\left(r_0/r_{\rm 0, i}\right)^{-n}$, then the total and time-dependent kinetic energy contained in the expanding gas is
\begin{equation}
\begin{split}
&E_{\rm kin} = 4\pi \int \frac{1}{2}v^2\rho r^2 dr \\
&=\frac{4\pi G\delta M \rho_{\rm 0, i}r_{\rm 0, i}^3}{3r_{\rm 0, i}}\tau^{\frac{2}{3}\left(2-n\right)}\int_0^{\min\left[\tau,\eta_{\rm sh}\right]}\left(\frac{df}{d\eta}\right)^2\eta^{\frac{2n}{3}-\frac{7}{3}}d\eta \label{Ekin} \\
&\equiv E_{\rm 0, i}\tau^{\frac{2}{3}\left(2-n\right)}\int_0^{\min\left[\tau,\eta_{\rm sh}\right]}\left(\frac{df}{d\eta}\right)^2\eta^{\frac{2n}{3}-\frac{7}{3}}d\eta.
\end{split}
\end{equation}
Here
\begin{equation}
E_{\rm 0, i} = \frac{4\pi G\delta M\rho_{\rm 0, i}r_{\rm 0, i}^3}{3 r_{\rm 0, i}}, \,\,\, \tau = \frac{\sqrt{G\delta M}t}{r_{\rm 0, i}^{3/2}},
\end{equation}
and the upper limit on the integral arises from the fact that the caustic forms in a dimensionless time $\eta_{\rm sh}$, implying that for times less than $\eta_{\rm sh}$ we integrate from the expanding inner edge, but for times later than $\eta_{\rm sh}$ we only integrate from the location of the caustic.

\begin{figure}[htbp] 
   \centering
   \includegraphics[width=0.47\textwidth]{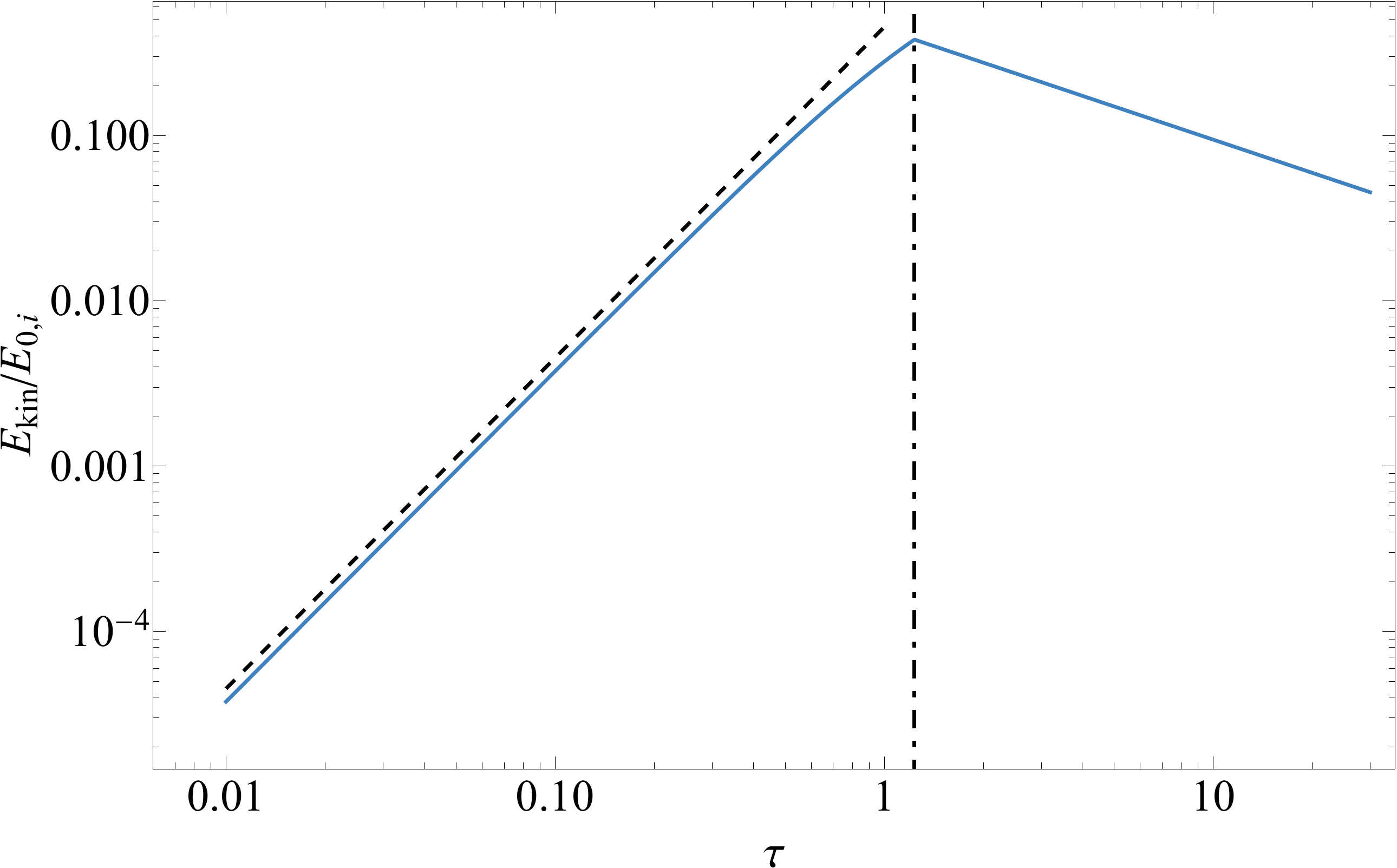} 
   \caption{The kinetic energy contained in the expanding gas as a function of time when the density of the ambient gas falls off as $\rho \propto r^{-2.5}$. The dashed line is $\propto \tau^2$, and is the scaling we expect in the limit that $\tau \ll \eta_{\rm sh}$, and the vertical, dot-dashed line coincides with $\tau = \eta_{\rm sh}$ and is where the kinetic energy is maximized. After $\tau = \eta_{\rm sh}$ the energy declines as $\propto \tau^{-1/3}$. }
   \label{fig:Ekin}
\end{figure}

Figure \ref{fig:Ekin} shows the kinetic energy contained in the gas, normalized by $E_{\rm 0, i}$, as a function of dimensionless time $\tau$ for $n = 2.5$. The dashed line is $\propto \tau^2$, and is what we expect based on the early-time scaling of the velocity as $v \simeq G\delta M t/r_0^2$. The dot-dashed, vertical line coincides with the time $\eta_{\rm sh}$, and is when the kinetic energy contained in the expanding gas is maximized at a value of $E_{\rm kin, max}/E_{\rm 0, i} \simeq 0.44$. For the YSG in \citet{fernandez18}, $\delta M = 0.12 M_{\odot}$ is the mass lost to neutrinos, $r_{\rm 0, i} \simeq 2.5\times 10^{11}$ cm is the base of the hydrogen envelope, and $\rho_{\rm 0, i} \simeq 5\times 10^{-4}$ g cm$^{-3}$ is the density at that radius, and we find $E_{\rm kin, max} \simeq 10^{45}$ erg; this energy is comparable to, but somewhat less than, the value inferred from the hydrodynamical simulations of \citet{fernandez18}, which implies that the energy is established at smaller radii. For times $\tau > \eta_{\rm sh}$, the energy contained in the dynamically expanding fluid declines as $\propto \tau^{-1/3}$. 

By considering only the dynamical response of the fluid to the mass loss, we see that a caustic forms at a location $\propto t^{2/3}$, and implies that a shock will form -- rendering gas pressure non-negligible. In the next section we show that, when the ambient density profile is a power-law with radius, there are self-similar solutions for the response of the gas at large radii that include the effects of pressure. 

\section{Pre-shock, self-similar Response to Neutrino Mass Loss}
\label{sec:ambient}
At the time of the neutrino-induced mass loss, we assume that there is at least some fraction of the stellar envelope that can be approximated as having a power-law density profile with power-law index $n$. At $t = 0$, which coincides with when the mass loss is assumed to occur instantaneously, we therefore have\footnote{In this and the next section we use an Eulerian formalism, thus treating the velocity as the dependent variable and radius as the independent variable, as this most readily facilitates the joining of the ambient solution onto a shock. We note, however, that it is possible to derive Lagrangian solutions for the response of the envelope by using the same assumption as in Equation \eqref{rss}. }
\begin{equation}
\rho(t = 0) = \rho_{\rm i}\left(\frac{r}{r_{\rm i}}\right)^{-n}. \label{rhoinit}
\end{equation}
Here $r$ is radial distance from the origin, while $\rho_{\rm i}$ and $r_{\rm i}$ are scale densities and radii, respectively. We also assume that the gas is initially in hydrostatic balance and that the mass interior to the envelope, which we denote $M$, dominates the mass of the envelope itself. The initial pressure profile of the envelope is therefore
\begin{equation}
p(t = 0) = \frac{1}{n+1}\frac{GM\rho_{\rm i}}{r}\left(\frac{r}{r_{\rm i}}\right)^{-n}. \label{pinit}
\end{equation}
The envelope responds hydrodynamically following the mass loss, and if at least some finite region of the envelope behaves as a smooth and ideal fluid, the gas obeys the inviscid fluid equations; in spherical symmetry (and accounting for the gravitational field of the mass $M$) and Eulerian form these equations are
\begin{equation}
\begin{split}
&\frac{\partial \rho}{\partial t}+\frac{1}{r^2}\frac{\partial}{\partial r}\left[r^2\rho v\right] = 0, \\ 
&\frac{\partial v}{\partial t}+v\frac{\partial v}{\partial r}+\frac{1}{\rho}\frac{\partial p}{\partial r} = -\frac{GM}{r^2}+\frac{G\delta M}{r^2}, \\ 
&\frac{\partial}{\partial t}\ln\left(\frac{p}{\rho^{\gamma}}\right)+v\frac{\partial}{\partial r}\ln\left(\frac{p}{\rho^{\gamma}}\right) = 0. \label{ent}
\end{split}
\end{equation}
These are the continuity, radial momentum, and entropy equations, respectively, where $v$ is the radial velocity and $\gamma$ is the adiabatic index of the gas. The mass lost to neutrinos is given by $\delta M$, and we have assumed that the mass is reduced by this amount in the sign of this additional term, i.e., $\delta M > 0$

Because the mass loss occurs instantaneously, the only relevant timescale is the dynamical time of the gas at radius $r$, $\tau_{\rm dyn} = r^{3/2}/\sqrt{GM}$, and there is no inherent radial scale (i.e., $r_{\rm i}$ in Equation \ref{rhoinit} is arbitrary). Given these observations, it is reasonable to expect that there exist self-similar solutions to Equations \eqref{ent} of the form
\begin{equation}
\begin{split}
v &= \sqrt{\frac{GM}{r}}f_{\rm e}(\eta), \quad \rho = \rho_{\rm i}\left(\frac{r}{r_{\rm i}}\right)^{-n}g_{\rm e}(\eta),\\
p &= \frac{1}{n+1}\frac{GM\rho_{\rm i}}{r}\left(\frac{r}{r_{\rm i}}\right)^{-n}h_{\rm e}(\eta), \label{ssdefs}
\end{split}
\end{equation}
where
\begin{equation}
\eta = \frac{t}{\tau_{\rm dyn}(r)} = \frac{t\sqrt{GM}}{r^{3/2}}
\end{equation}
and $f_{\rm e}$, $g_{\rm e}$, and $h_{\rm e}$ are functions that satisfy the initial conditions (from hydrostatic balance)
\begin{equation}
f_{\rm e}(0) = 0, \,\,\, g_{\rm e}(0) = h_{\rm e}(0) = 1. \label{inits}
\end{equation}
Inserting Equations \eqref{ssdefs} into the fluid equations and performing some algebraic rearrangements yields the three following ordinary differential equations for the functions $f_{\rm e}$, $g_{\rm e}$, and $h_{\rm e}$:
\begin{equation}
\frac{\partial g_{\rm e}}{\partial \eta}+\left(\frac{3}{2}-n\right)f_{\rm e}g_{\rm e}-\frac{3}{2}\eta\frac{\partial}{\partial \eta}\left[f_{\rm e}g_{\rm e}\right] = 0,
\end{equation}
\begin{multline}
\frac{\partial f_{\rm e}}{\partial \eta}-\frac{1}{2}f_{\rm e}\left(f_{\rm e}+3\eta\frac{\partial f_{\rm e}}{\partial \eta}\right) \\ 
-\frac{1}{n+1}\frac{1}{g_{\rm e}}\left(\left(n+1\right)h_{\rm e}+\frac{3}{2}\eta\frac{\partial h_{\rm e}}{\partial \eta}\right) 
= -1+\frac{\delta M}{M},
\end{multline}
\begin{equation}
\frac{\partial}{\partial \eta}\ln\left[\frac{h_{\rm e}}{g_{\rm e}^{\gamma}}\right]+f_{\rm e}\left(n\gamma-n-1-\frac{3}{2}\eta\frac{\partial}{\partial \eta}\ln\left[\frac{h_{\rm e}}{g_{\rm e}^{\gamma}}\right]\right) = 0. \label{sseqs}
\end{equation}
Solutions to these equations describe the instantaneous response of the envelope at large radii, i.e., the initial conditions \eqref{inits} also apply in the limit that $r \rightarrow \infty$. 

\begin{figure*}[t] 
   \centering
   \includegraphics[width=0.495\textwidth]{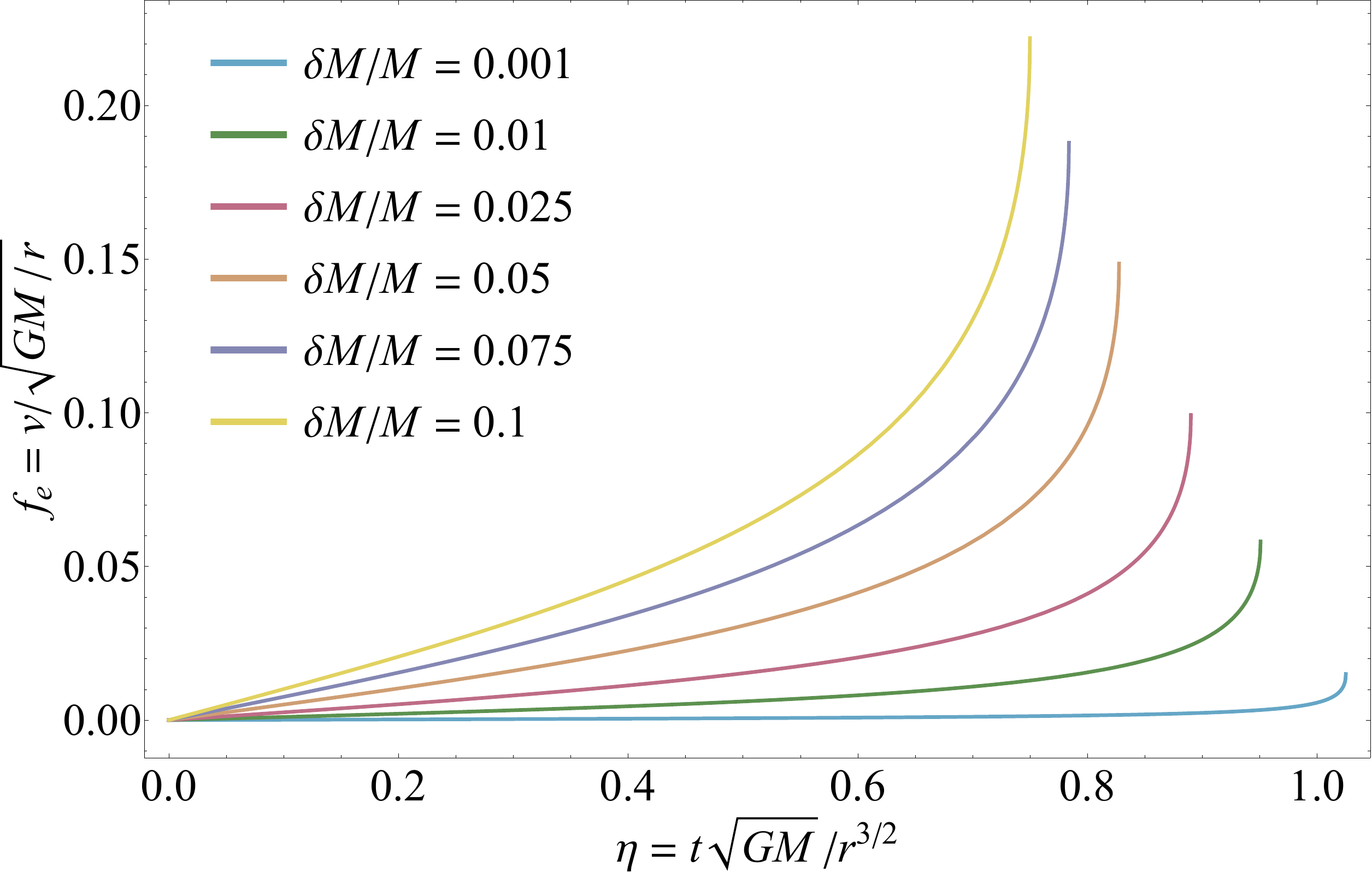} 
 \includegraphics[width=0.495\textwidth]{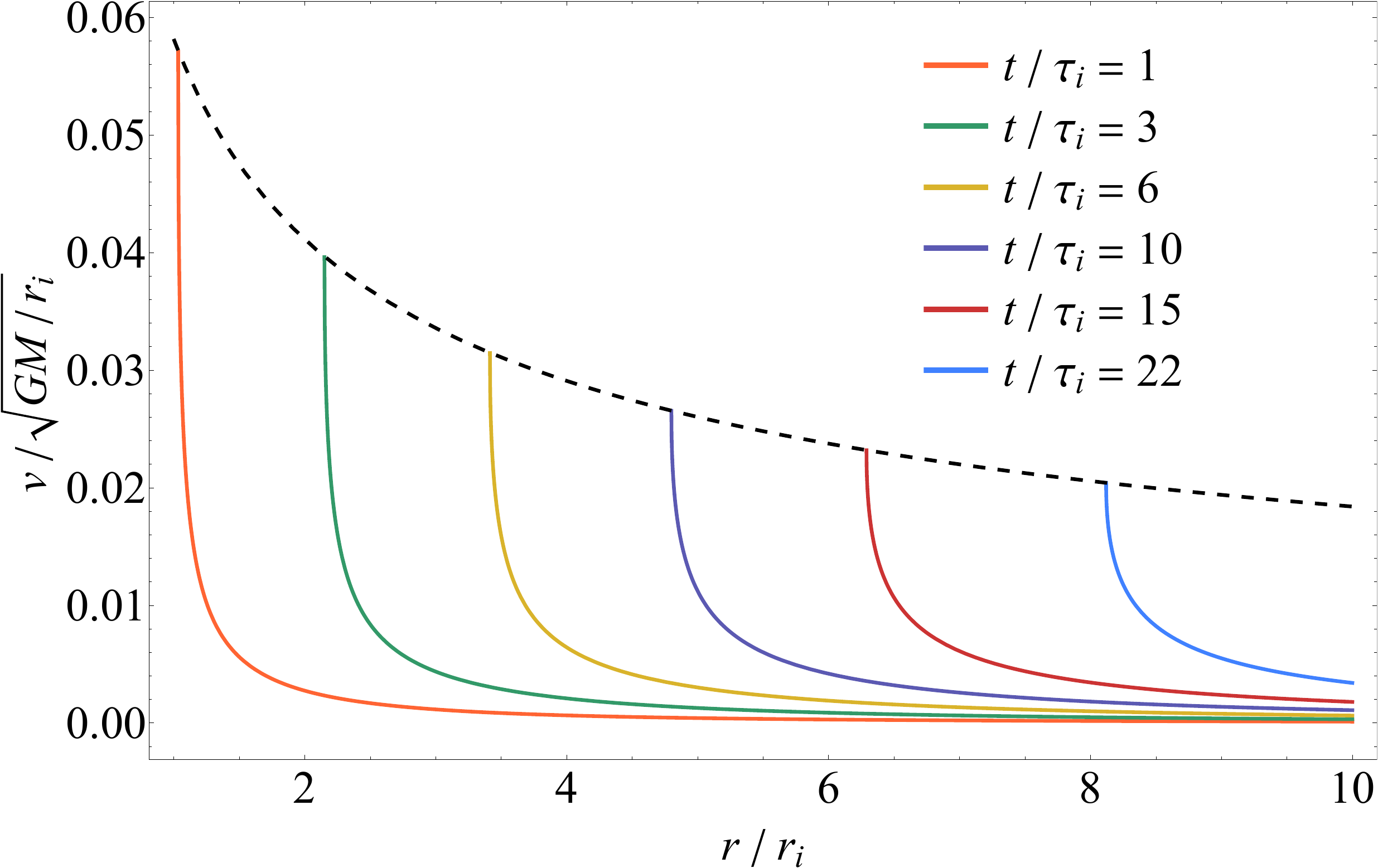} 
   \caption{Left: The self-similar velocity $f_{\rm e}$ (i.e., $v/\sqrt{GM/r}$) as a function of the self-similar variable $\eta = t/\tau_{\rm dyn} = t\sqrt{GM}/r^{3/2}$ for $n = 2.5$, $\gamma = 1.4$, and the fractional mass losses in the legend. For small $\eta$ (early times and/or large radii) the solution is approximately $f_{\rm e} \simeq \eta\delta M/M$, but for sufficiently large $\eta$ (i.e., at small enough radii for a given time) the solution becomes nonlinear and terminates in a sonic point. Right: The velocity normalized by $\sqrt{GM/r_{\rm i}}$ as a function of radius normalized by $r_{\rm i}$ for the times shown in the legend, where $\tau_{\rm i} = r_{\rm i}^{3/2}/\sqrt{GM}$ is the dynamical time at $r_{\rm i}$. This shows that an outward velocity is generated in response to the mass loss, and this velocity decays as $\propto r^{-2}$ at large radii, but ends in a sonic point at small radii. The black, dashed line is $\propto 1/\sqrt{r}$, and joins the velocity at the sonic point at each time. }
   \label{fig:v_vesc}
\end{figure*}

While the exact solutions to \eqref{sseqs} must be determined numerically, they can be written as power series in $\eta$, the lowest-order terms of which are 
\begin{equation}
f_{\rm e} = \frac{\delta M}{M}\eta = \frac{G\delta M t}{r^2}, \quad g_{\rm e} = h_{\rm e} = 1. \label{solapprox}
\end{equation}
This result -- that the initial response of the gas is purely dynamical with no change to the density and pressure -- agrees with the analysis in \citet{coughlin18}, who solved the linearized (in $\delta M/M$) equations with an eigenvalue approach to understand the formation and steepening of the sound wave (launched during the mass loss and near the radius where the dynamical and neutrino mass-loss timescales are comparable). We therefore expect that the values of $n$ and $\gamma$ will only affect the solution for relatively large values of $\eta = t\sqrt{GM}/r^{3/2}$, i.e., for late times after the mass loss and/or small radii. The solutions (both those obtained from Equations \ref{sseqs} and the approximate solutions \ref{solapprox}) also depend only on the {relative} mass loss, $\delta M/M$. In what follows our fiducial case will be $n = 2.5$, $\gamma = 1.4$, and $\delta M/M = 0.01$, as these values are appropriate to the YSG from \citet{fernandez18} (at the time of core-collapse the YSG has a mass of $M = 11.1 M_{\odot}$ and $\delta M = 0.12 M_{\odot}$, or $\delta M/M \simeq 0.0108$). 

The left panel of Figure \ref{fig:v_vesc} shows the self-similar velocity $f_{\rm e}$, i.e., the velocity relative to $\sqrt{GM/r}$, as a function of $\eta = t\sqrt{GM}/r^{3/2}$, for $n = 2.5$, $\gamma = 1.4$, and the relative mass losses shown in the legend. For small $\eta$ (early times or large radii) the velocity increases as $\delta M \eta/M$. As $\eta$ increases (so for a fixed radius as time increases, or for a fixed time as radius decreases), the solution displays an increasing degree of nonlinearity, ultimately terminating at a sonic point. The right panel shows the velocity, normalized by $\sqrt{GM/r_{\rm i}}$, as a function of radius normalized by $r_{\rm i}$, for $n = 2.5$, $\gamma = 1.4$, and $\delta M/M = 0.01$, at the times shown in the legend, where $\tau_{\rm i} = r_{\rm i}^{3/2}/\sqrt{GM}$ is the dynamical time at the scale radius. For large radii and early times, the velocity decays approximately as $\propto 1/r^2$, which is in agreement with expectations from the analytical (leading-order; Equation \ref{solapprox}) solution. At small radii, however, the solution ends in a sonic point, and the black, dashed line -- which scales as $\propto r^{-1/2}$ -- joins the velocity at the sonic point at each time. 

\begin{figure}[h] 
   \centering
   \includegraphics[width=0.475\textwidth]{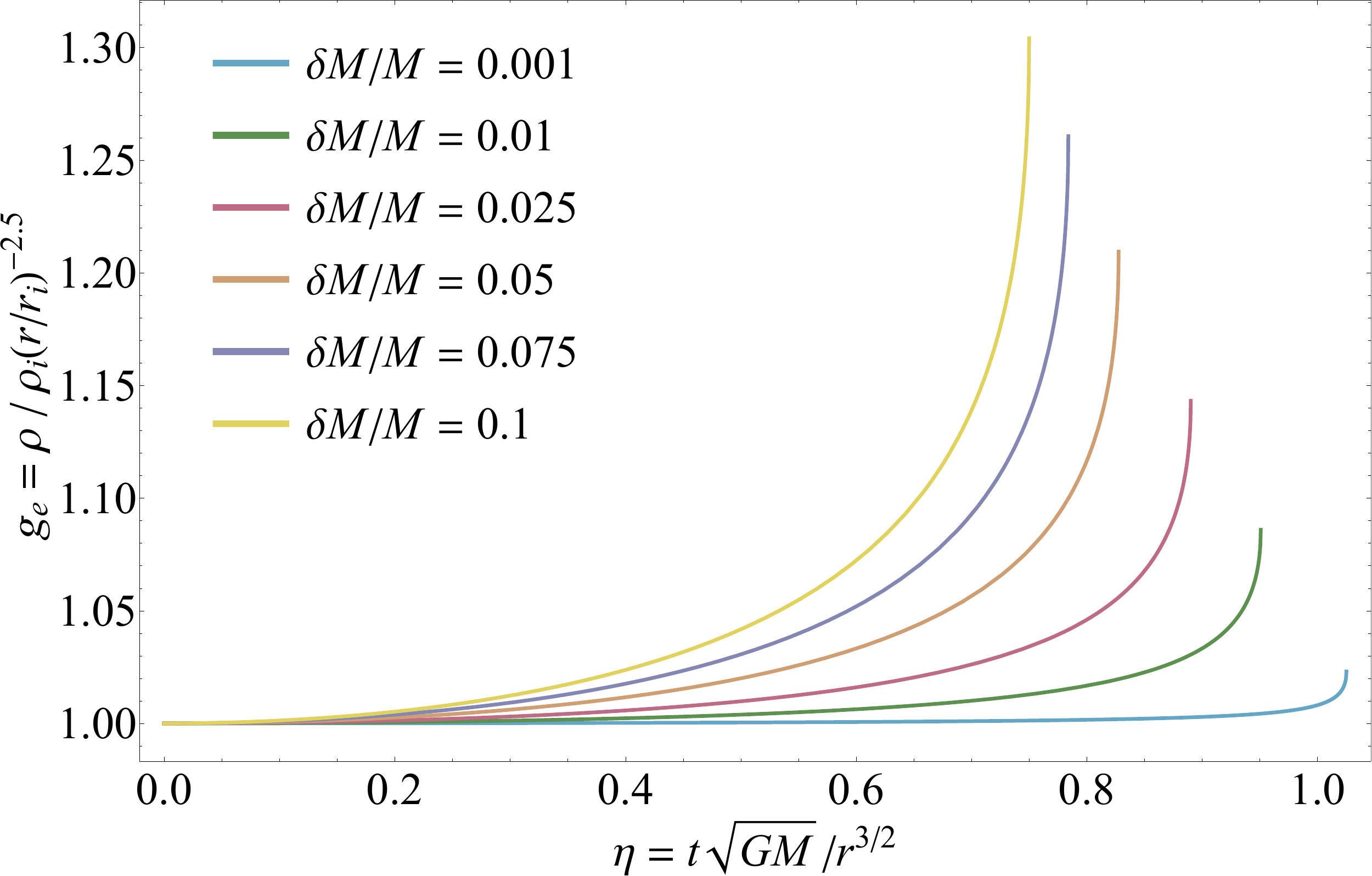} 
   \caption{Left: The self-similar density $g_{\rm e}$ (i.e., the density normalized by $\rho_{\rm i}\left(r/r_{\rm i}\right)^{-n}$) as a function of the self-similar variable $\eta = t/\tau_{\rm dyn} = t\sqrt{GM}/r^{3/2}$ for $n = 2.5$, $\gamma = 1.4$, and the $\delta M/M$ in the legend. The solutions are indistinguishable for small $\eta$ and equal to $\simeq 1$, which is consistent with the perturbative limit, and only once the solution approaches the sonic point are there noticeable deviations from unity.}
   \label{fig:ge_n2p5}
\end{figure}

\begin{figure}[h] 
   \centering
   \includegraphics[width=0.475\textwidth]{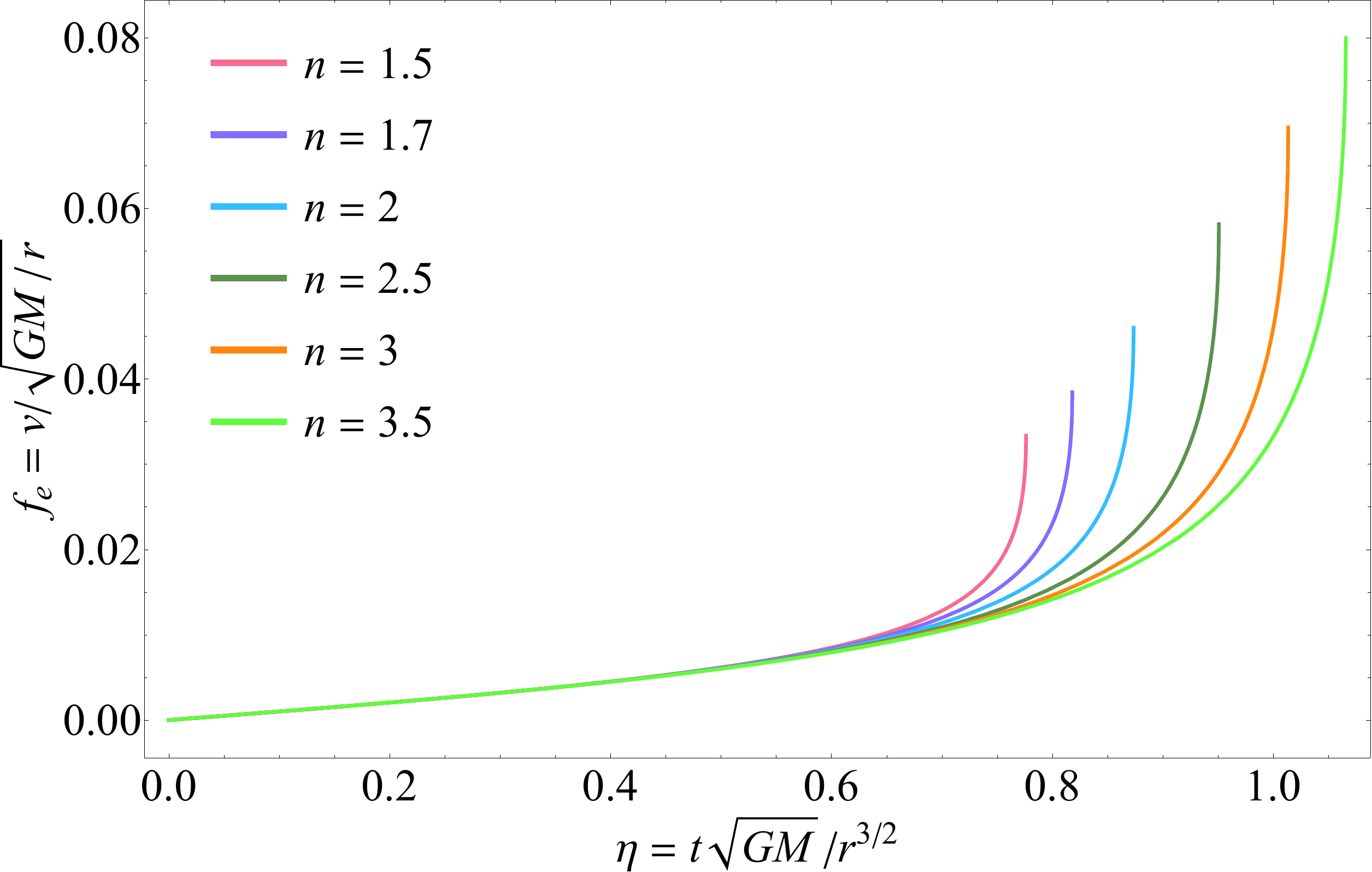} 
   \caption{Left: The self-similar velocity $f_{\rm e}$ (i.e., $v/\sqrt{GM/r}$) as a function of the self-similar variable $\eta = t/\tau_{\rm dyn} = t\sqrt{GM}/r^{3/2}$ for $\delta M/M = 0.01$, $\gamma = 1+1/n$, and the $n$ (power-law index of the ambient medium) shown in the legend. The solutions are indistinguishable for small $\eta$, which is consistent with the perturbative limit.}
   \label{fig:fe_of_n}
\end{figure}

Figure \ref{fig:ge_n2p5} shows the self-similar density $g_{\rm e}$ (i.e., the density normalized by $\rho_{\rm i}\left(r/r_{\rm i}\right)^{-n}$) for $n = 2.5$, $\gamma = 1.4$, and the fractional mass losses shown in the legend. As expected from the leading-order solution, for small $\eta$ we have $g_{\rm e} \simeq 1$, and hence the density is roughly unaffected by the change in the mass loss. The solution for the self-similar pressure, $h_{\rm e}$, is qualitatively similar, and hence we do not plot it. Figure \ref{fig:fe_of_n} illustrates the self-similar velocity for $\delta M/M = 0.01$ and the $n$ shown in the legend; here we set $\gamma = 1+1/n$, i.e., the envelope is polytropic, which is approximately valid for convective supergiant envelopes \citep{coughlin18b}. The solutions are indistinguishable at small radii, and satisfy $f_{\rm e} \simeq \delta M/M\times \eta$. The trend with the power-law index is that the sonic point is pushed to a larger value of the self-similar variable as $n$ increases. 

\begin{figure*}[t] 
   \centering
   \includegraphics[width=0.495\textwidth]{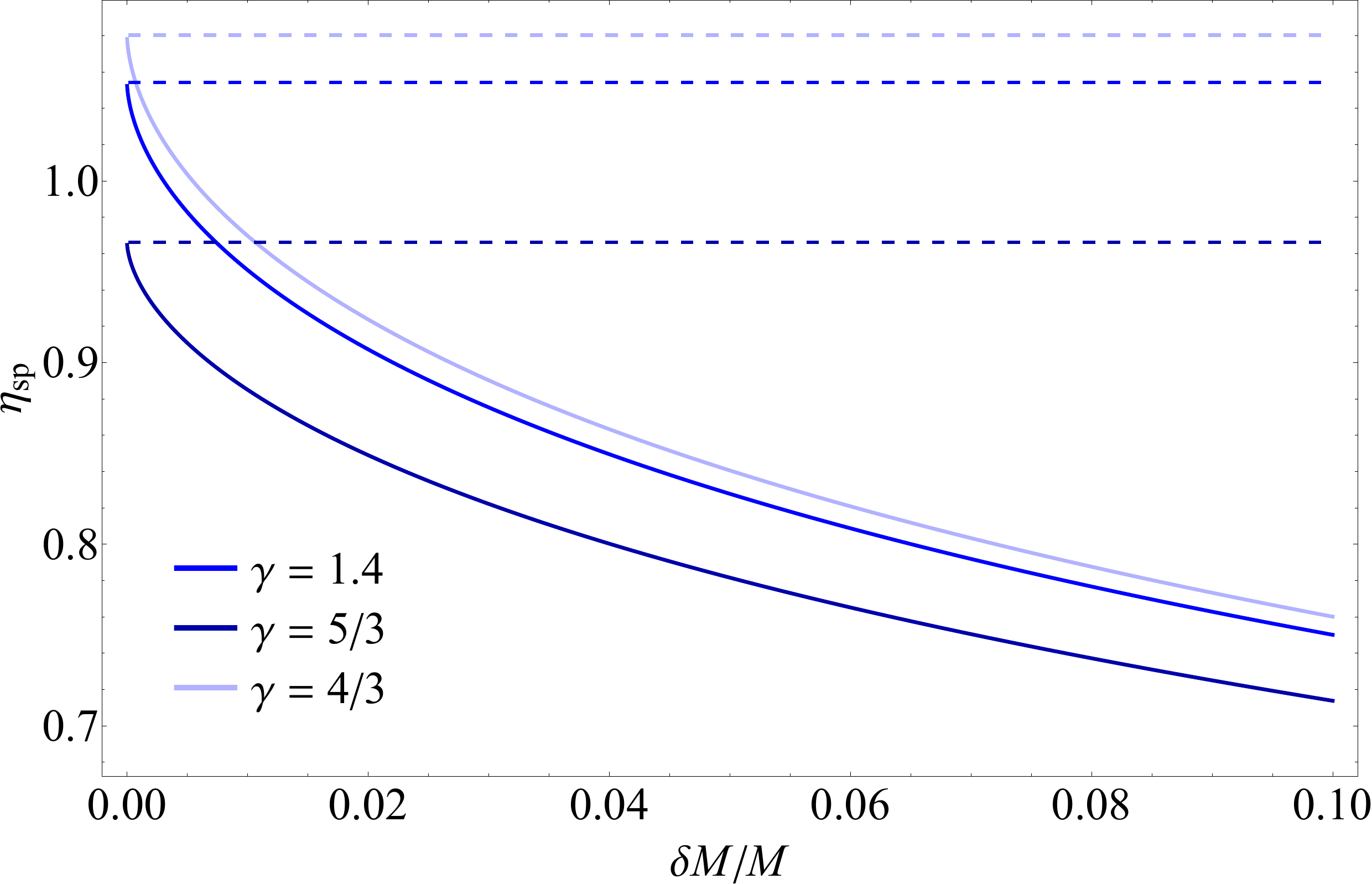}
   \includegraphics[width=0.495\textwidth]{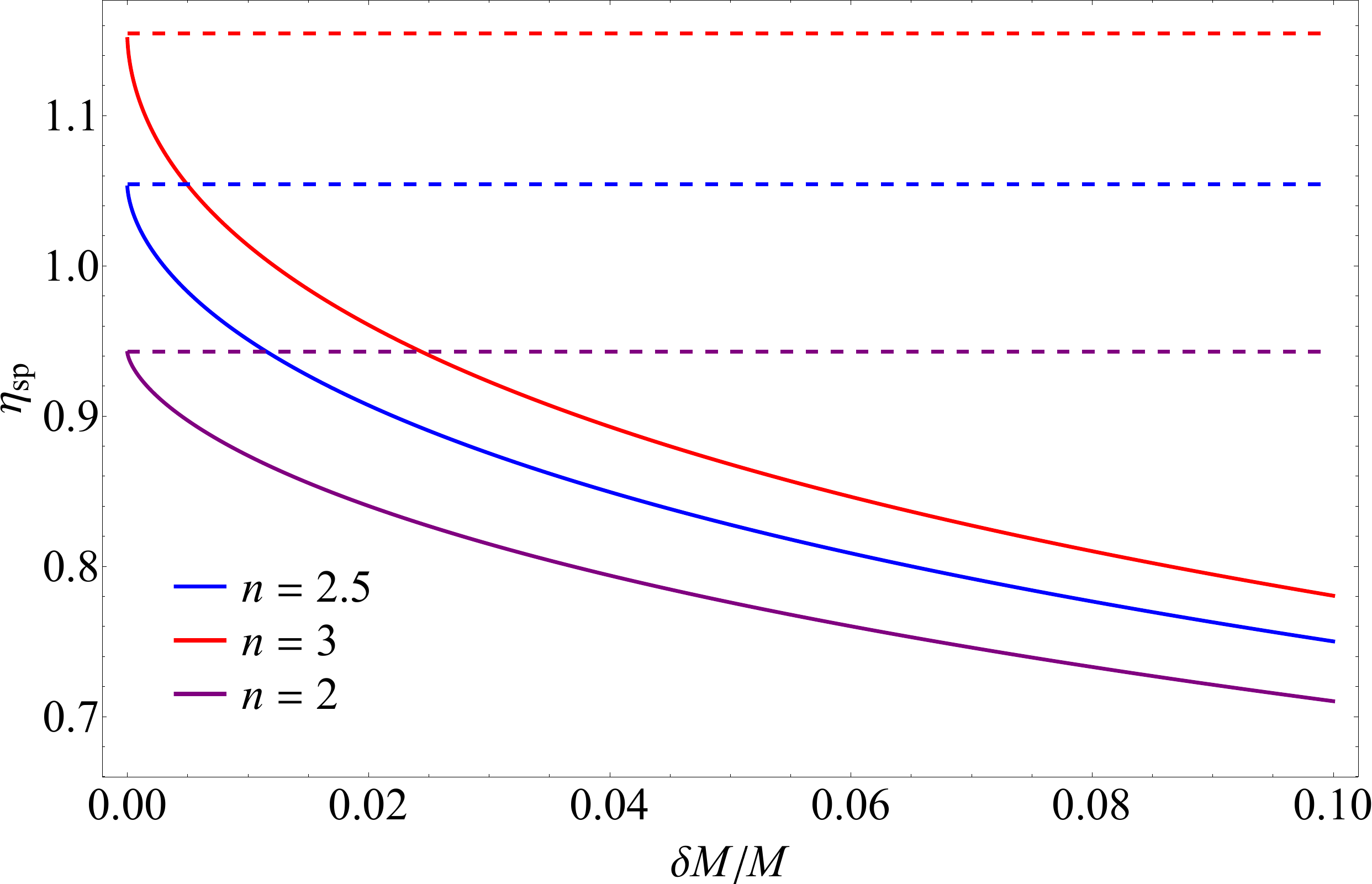}
   \caption{Left: The dimensionless sonic point $\eta_{\rm sp}$ as a function of $\delta M/M$ for $n = 2.5$ and the $\gamma$ in the legend. The horizontal, dashed lines give the value of $\eta_{\rm sp}$ appropriate to the sound speed of the envelope, being $2\sqrt{(n+1)/\gamma}/3$, which shows that $\eta_{\rm sp}$ approaches this value in the limit that $\delta M/M \rightarrow 0$ (as it must), and that the sonic radius moves out at a speed that is always greater than the hydrostatic sound speed for any finite $\delta M/M$. Right: Same as the left panel, but with $\gamma = 1+1/n$ and the values of $n$ in the legend. }
   \label{fig:etasp}
\end{figure*}

Figures \ref{fig:v_vesc} -- \ref{fig:fe_of_n} demonstrate that the solutions for the expansion of the envelope are characterized by the existence of a sonic point, which we define as $\eta_{\rm sp}$, implying that the sonic radius $R_{\rm sp}$ expands as
\begin{equation}
R_{\rm sp} = \left(\frac{\sqrt{GM}t}{\eta_{\rm sp}}\right)^{2/3}, \,\,\, V_{\rm sp} = \frac{dR_{\rm sp}}{dt} = \frac{2}{3\eta_{\rm sp}}\sqrt{\frac{GM}{R_{\rm sp}}} \label{Rsp}
\end{equation}
The sound speed of the hydrostatic medium is given by $c_{\rm s} = \sqrt{\gamma/(n+1)}\sqrt{GM/r}$, and because we expect $V_{\rm sp} > c_{\rm s}$ (i.e., the outward motion of the gas means that the sonic point must expand faster than the hydrostatic value), this implies $\eta_{\rm sp} \le 2\sqrt{(n+1)/\gamma}/3$. The left panel of Figure \ref{fig:etasp} shows $\eta_{\rm sp}$ as a function of $\delta M/M$ for $n = 2.5$ and the values of $\gamma$ in the legend, and the right panel shows $\eta_{\rm sp}$ as a function of $\delta M/M$ for $\gamma = 1+1/n$ and the values of $n$ in the legend. In each plot the dashed lines give the limiting value $2\sqrt{(n+1)/\gamma}/3$, and demonstrate that $\eta_{\rm sp}$ is always less than this value, implying that the sonic radius in the expanding flow moves out at a speed that is faster than the hydrostatic sound speed, and approaches this value in the limit that $\delta M/M \rightarrow 0$. The right panel is the same as the left panel, but in this case we let $\gamma = 1+1/n$ and the value of $n$ appropriate to each curve is shown in the legend. 

The fact that these solutions terminate in a sonic point implies that there must be a shock that moves out at some time-dependent radius $R_{\rm sh}(t) > R_{\rm sp}(t)$ that connects the inner region to the expanding envelope, and in the next section we show that -- for values of $\delta M/M$ below a critical value -- there are self-similar, weak shock solutions that satisfy $R_{\rm sh}\propto t^{2/3}$ and achieve this. Before doing so, however, it is useful to consider the total kinetic energy contained in the expanding envelope. To do so, we assume that the scale radius $r_{\rm i}$ coincides with the location of the rarefaction wave at the time the mass loss occurs, and any mass interior to this radius is accreted onto the black hole (and is not contained in the expanding envelope). Then the total energy is
\begin{equation*}
E_{\rm kin}(t) = 4\pi \int_{\max\left[r_{\rm i}, R_{\rm sp}(t)\right]}^{\infty}\frac{1}{2}v^2\rho r^2dr 
\end{equation*}
\begin{equation}
=E_{\rm i}\tau^{\frac{2}{3}\left(2-n\right)}
\times \int_0^{\min\left[\tau\left(\frac{r[\tau]}{r_{\rm i}}\right)^{-3/2},\eta_{\rm sp}\right]}f_{\rm e}^2g_{\rm e}\eta^{\frac{2n}{3}-\frac{7}{3}}d\eta, \label{KEint}
\end{equation}
where
\begin{equation}
E_{\rm i} \equiv \frac{4\pi}{3}\frac{GM}{r_{\rm i}}\rho_{\rm i}r_{\rm i}^3, \quad \tau = \frac{\sqrt{GM}t}{r_{\rm i}^{3/2}}.
\end{equation}
Equation \eqref{KEint} bears similarity to Equation \eqref{Ekin}, but now the definitions of $E_{\rm i}$ and $\tau$ involve not the mass lost, but the background mass $M$. The scaling with the fractional mass loss, $\delta M/M$, is also contained entirely within the integral. As for Equation \eqref{Ekin}, the upper limit of integration implies that we integrate to the expanding inner edge of the envelope until the sonic point is contained within the flow, after which time we integrate to $\eta_{\rm sp}$. 

\begin{figure*}[t] 
   \centering
   \includegraphics[width=0.495\textwidth]{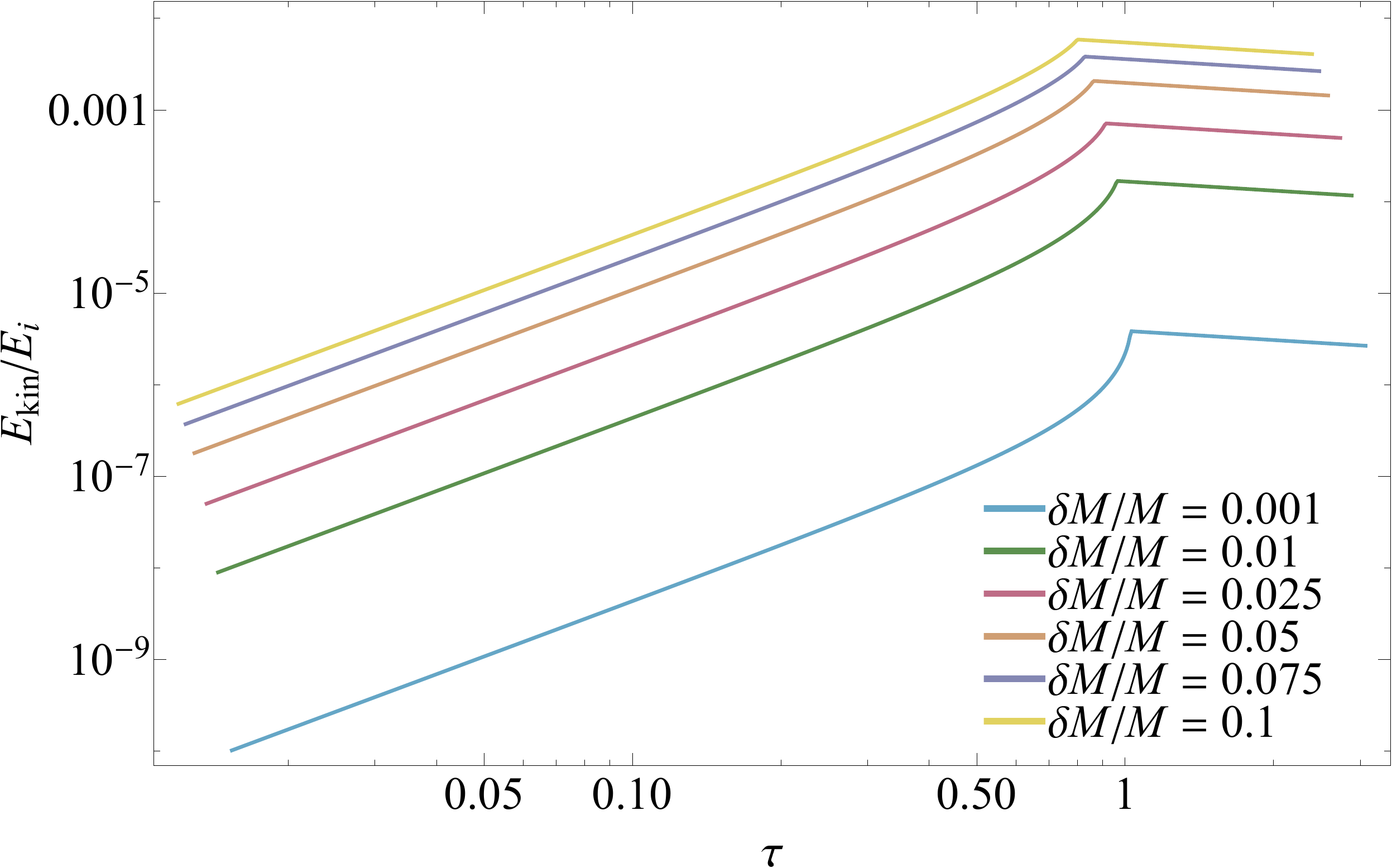} 
   \includegraphics[width=0.495\textwidth]{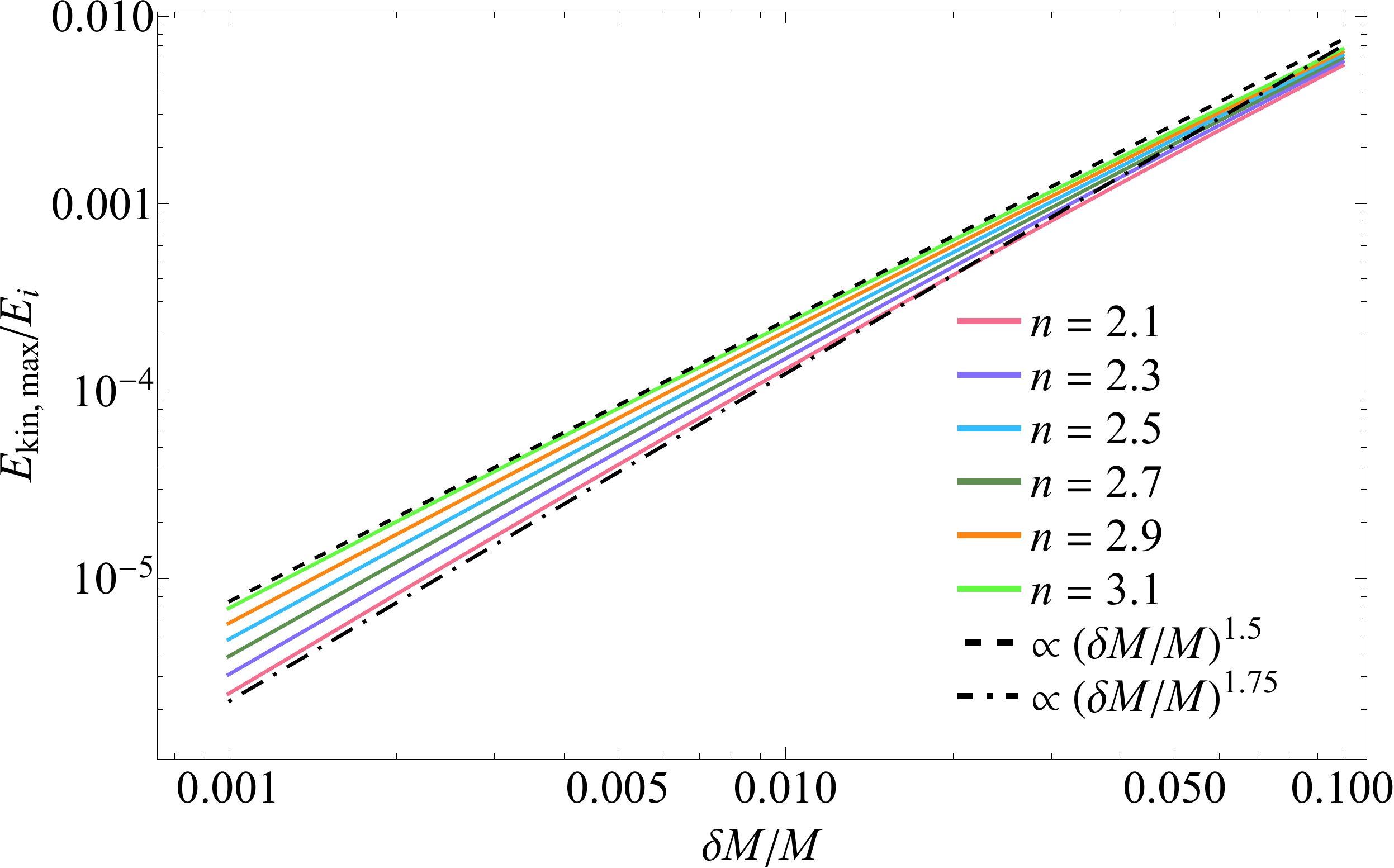} 
   \caption{Left: The kinetic energy contained in the expanding gas as a function of dimensionless time $\tau$ for $n = 2.5$, $\gamma = 1.4$, and the fractional mass losses shown in the legend. The scaling at small $\tau$ is $\propto \tau^2$, and each curve reaches a peak at a time of $\tau = \eta_{\rm sp}$, which is when the sonic point forms at the inner edge of the flow. Each curve subsequently decays as $\propto \tau^{-1/3}$. Right: The maximum kinetic energy (i.e., corresponding to the peaks in the left panel) as a function of the relative mass loss for the density power-law indices shown in the legend and $\gamma = 1+1/n$. The dashed and dot-dashed curves show power-law scalings with the mass loss that are approximately upheld for $n = 2.1$ and $n = 3.1$ at small $\delta M/M$.}
   \label{fig:Ekin_dM_n}
\end{figure*}

The left panel of Figure \ref{fig:Ekin_dM_n} shows the kinetic energy normalized by $E_{\rm i}$ as a function of $\tau$ for the mass losses shown in the legend, and here we set $n = 2.5$ and $\gamma = 1.4$. The kinetic energy initially increases as $\propto \tau^2$ for $\tau \simeq 0$, reaches a peak at $\tau = \eta_{\rm sp}$, and subsequently declines as $\propto \tau^{-1/3}$. As the mass loss increases, the maximum-attained energy correspondingly increases, and to highlight this behavior the right panel of this figure illustrates the maximum kinetic energy (again, normalized by $E_{\rm i}$) as a function of $\delta M/M$ for the ambient power-law indices in the legend and $\gamma = 1+1/n$. 

The dashed and dot-dashed curves in the right panel of Figure \ref{fig:Ekin_dM_n} show that there are approximate, power-law scalings of the maximum energy in the expanding gas at small $\delta M/M$. \citet{coughlin18} predicted from the approximate solution for $v \propto \delta M$ that the energy should scale as $\delta M^2$, while the purely dynamic solutions in Section \ref{sec:basic} suggest that the maximum energy should satisfy $E_{\rm kin, max} \propto \delta M$. We see that the scaling from the self-similar solutions is, at least for $\delta M/M \lesssim few\times 0.01$, between these two limits. The reason for this is that the analysis in \citet{coughlin18} did not account for the fact that the time at which the shock would form in the flow would depend on $\delta M/M$, i.e., they assumed the $v \propto \delta M$ scaling would hold until a pre-determined time dependent solely on the progenitor (their Equation 4). The analysis in Section \ref{sec:basic}, on the other hand, did not account for the finite sound speed in the ambient gas and the fact that the sonic point forms at a time significantly earlier than the caustic. In particular, the radius of the caustic expands as $\propto \delta M^{1/3}t^{2/3}$, while the sonic point is roughly independent of $\delta M$ for $\delta M/M \ll 1$ (see Figure \ref{fig:etasp}). Consequently, if we let $n = 2.5$, $\gamma = 1.4$, and $\delta M/M = 0.01$, we find $E_{\rm kin, max}/E_{\rm i} \simeq 1.68\times 10^{-4}$, and using $M = 11.2 M_{\odot}$, $r_{\rm 0, i} \simeq 2.5\times 10^{11}$ cm, and $\rho_{\rm 0, i} \simeq 5\times 10^{-4}$ g cm$^{-3}$ (the values appropriate to the YSG in \citealt{fernandez18}), then we find $E_{\rm kin, max} \simeq 3.3\times 10^{43}$ erg -- a factor of $\sim 30$ smaller than the value inferred from assuming a purely dynamical response of the envelope. 

The energy derived here assumes that the flow extends to the sonic point, but there must be a shock at larger radii that joins the solution onto the inner flow. In the next section we show that there are self-similar solutions for a weak shock (and the post-shock flow), provided that the mass loss is below a critical fraction, that join onto the expanding envelope and yield accretion -- as well as outward-moving gas -- onto the black hole. 

\section{Weak-Shock, Self-similar solutions}
\label{sec:shock}
To join the expanding envelope solutions derived in the previous section onto the inner region that is accreting onto the newly formed black hole, there must be a shock that moves out at a time-dependent position $R_{\rm sh}(t)$, and that position must satisfy $R_{\rm sh}(t) \ge R_{\rm sp}(t)$, where $R_{\rm sp}(t)$ is the sonic radius that characterizes the expanding envelope. From Section \ref{sec:basic} we saw that the purely dynamical response of the gas results in the formation of a caustic that expands outward with time as $\propto t^{2/3}$, and from the preceding section the sonic radius also satisfies $R_{\rm sp} \propto t^{2/3}$. Both of these results suggest that there may be self-similar solutions to the fluid equations for which the shock position satisfies 
\begin{equation}
\frac{\sqrt{GM}t}{R_{\rm sh}^{3/2}} = \eta_{\rm sh}, \,\,\, V_{\rm sh} = \frac{2}{3\eta_{\rm sh}}\sqrt{\frac{GM}{R_{\rm sh}}}, \,\,\, \eta_{\rm sh} < \eta_{\rm sp}, \label{Rsh}
\end{equation}
with $\eta_{\rm sh}$ a constant; the inequality $\eta_{\rm sh} < \eta_{\rm sp}$ ensures that the shock position leads the location of the sonic point. Then we write the post-shock fluid velocity, density, and pressure as
\begin{equation}
\begin{split}
v &= V_{\rm sh}\left(t\right) f_{\rm s}(\xi), \quad \rho = \rho_{\rm i}\left(\frac{R_{\rm sh}\left[t\right]}{r_{\rm i}}\right)^{-n}g_{\rm s}(\xi)\\
p &= \rho_{\rm i}\left(\frac{R_{\rm sh}\left[t\right]}{r_{\rm i}}\right)^{-n}V_{\rm sh}^2 h_{\rm s}(\xi), \quad \xi = \frac{r}{R_{\rm sh}\left(t\right)}.
\end{split}
\end{equation}
Inserting these into the fluid equations then yields, upon using Equation \eqref{Rsh} and making some trivial algebraic rearrangements, 
\begin{equation}
-ng_{\rm s}-\xi\frac{dg_{\rm s}}{d\xi}+\frac{1}{\xi^2}\frac{d}{d\xi}\left[\xi^2f_{\rm s}g_{\rm s}\right] = 0, \label{ss1}
\end{equation}
\begin{equation}
-\frac{1}{2}f_{\rm s}+\left(f_{\rm s}-\xi\right)\frac{df_{\rm s}}{d\xi}+\frac{1}{g_{\rm s}}\frac{dh_{\rm s}}{d\xi} = -\frac{9}{4}\eta_{\rm s}^2\frac{1}{\xi^2}\left(1-\frac{\delta M}{M}\right), \label{ss2}
\end{equation}
\begin{equation}
n\gamma-n-1+\left(f_{\rm s}-\xi\right)\frac{d}{d\xi}\ln\left(\frac{h_{\rm s}}{g_{\rm s}^{\gamma}}\right) = 0. \label{ss3}
\end{equation}
The fluid variables satisfy the jump conditions at the shock, and we assume that the adiabatic index remains unchanged across the shock, which is valid if the shock is not too strong (the case under consideration here). Because the jump conditions retain a non-zero ambient velocity and pressure, which are given by the solutions derived in the previous section evaluated at $\eta_{\rm sh}$, the expressions for $f_{\rm s}(1)$, $g_{\rm s}(1)$, and $h_{\rm s}(1)$ (i.e., at $r = R_{\rm sh}$) are lengthy. To retain the readability of the paper, we place the boundary conditions in Appendix \ref{sec:bcs} for the interested reader. 

With the jump conditions, Equations \eqref{ss1} -- \eqref{ss3} can be integrated numerically inward from $\xi = 1$ once we specify the value of $\eta_{\rm sh}$, which in general is not constrained by these equations or the boundary conditions. However, if we require that the solutions accrete onto the newly formed black hole at the origin, then we find for sufficiently small fractional mass losses that there are {two solutions} that smoothly pass through a sonic point in the interior of the flow (known as a type-II similarity solution; \citealt{sedov59}) and yield accretion. We denote the solution with the larger Mach number by $f_{\rm s, s}$ (i..e, the ``strong'' solution), and that with the smaller Mach number $f_{\rm s, w}$ (i.e., the ``weak'' solution), and similarly for the functions $g_{\rm s}$ and $h_{\rm s}$ and for the shock parameter $\eta_{\rm sh}$. (As we also discuss in Appendix \ref{sec:settling}, there is a distinct class of solution that has zero mass flux at the origin, is causally connected everywhere, and the gas ``settles'' onto the compact object such that the velocity approaches zero as $r\rightarrow 0$.)

\begin{figure}[h] 
   \centering
   \includegraphics[width=0.475\textwidth]{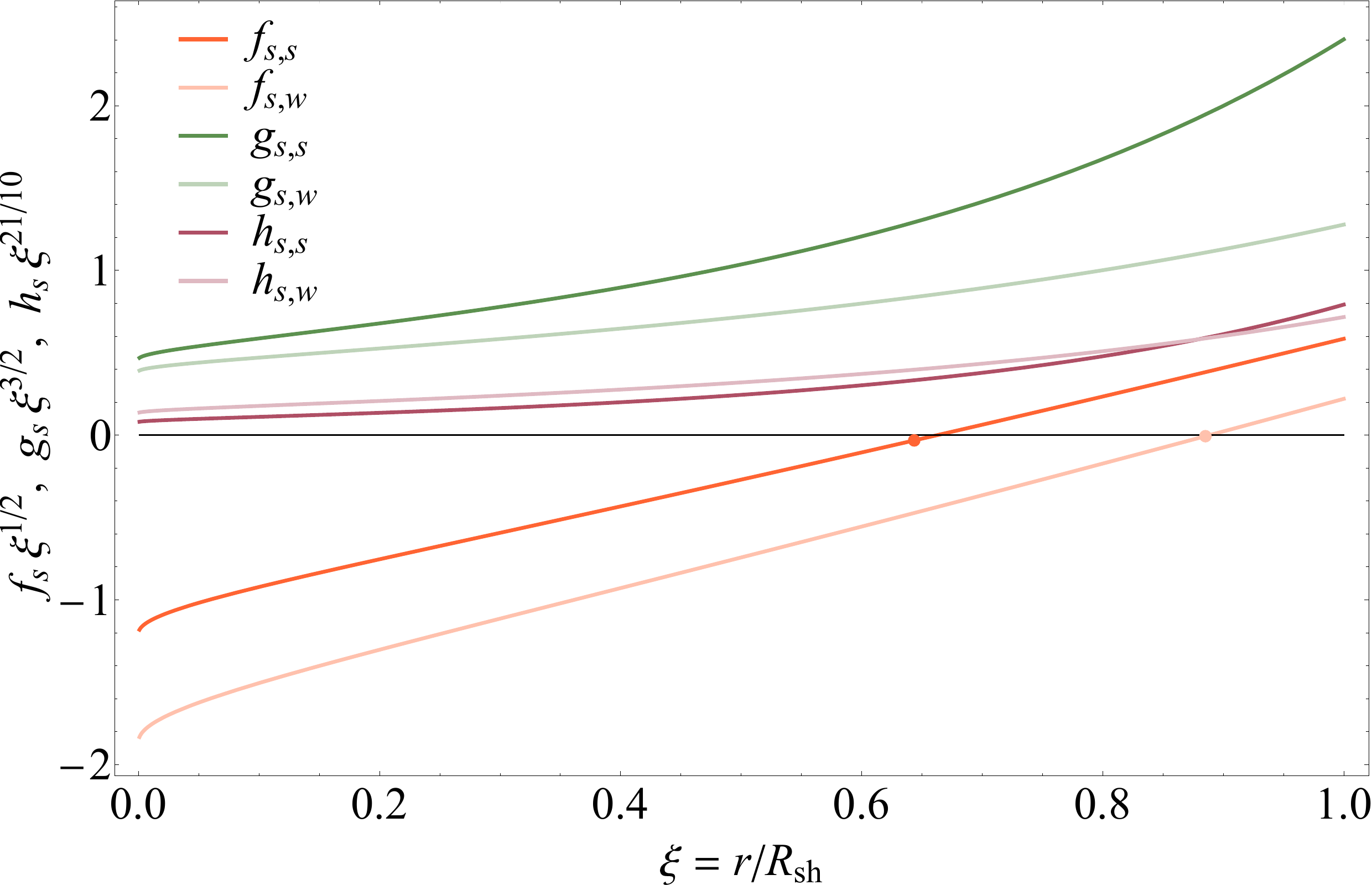} 
    \caption{The post-shock self-similar velocity, density, and pressure, normalized by $\xi^{-1/2}$, $\xi^{-3/2}$, and $\xi^{-21/10}$, respectively, as a function of the self-similar variable $\xi = r/R_{\rm sh}$, for $n = 2.5$, $\gamma = 1.4$, and $\delta M/M = 0.01$. The dark (light) curves show the strong-shock (weak-shock) solution that matches onto the expanding envelope, and the points show the location of the sonic radius within the flow, which is close to the stagnation radius where $v = 0$.}
   \label{fig:fs_gs}
\end{figure}

Figure \ref{fig:fs_gs} shows the weak and strong-shock solutions for $n = 2.5$, $\gamma = 1.4$, and $\delta M/M = 0.01$; these solutions have $\eta_{\rm sh, s} = 0.644$ and $\eta_{\rm sh, w} = 0.885$, which correspond to Mach numbers $\mathscr{M}_{\rm sh, s} = 1.83$ and $\mathscr{M}_{\rm sh, w} = 1.19$. The points on the velocity profiles illustrate the location of the sonic point for each solution, which are approximately coincident with where the velocity equals zero. Near the shock the velocity is positive, indicating the outward motion of the fluid as it is hit by the shock, but after receding through the sonic point the gas falls back onto the compact object at the origin. These solutions, similarly to those described in \citet{coughlin18b}, therefore simultaneously result in the outward motion of the gas near the shock and in fallback accretion onto the newly formed black hole. In both cases the velocity and density approach their freefall scalings near the origin, while the pressure approaches the adiabatic scaling, i.e., $h_{\rm s} \propto g_{\rm s}^{\gamma} \propto r^{-21/10}$. 

\begin{figure}[h] 
   \centering
   \includegraphics[width=0.475\textwidth]{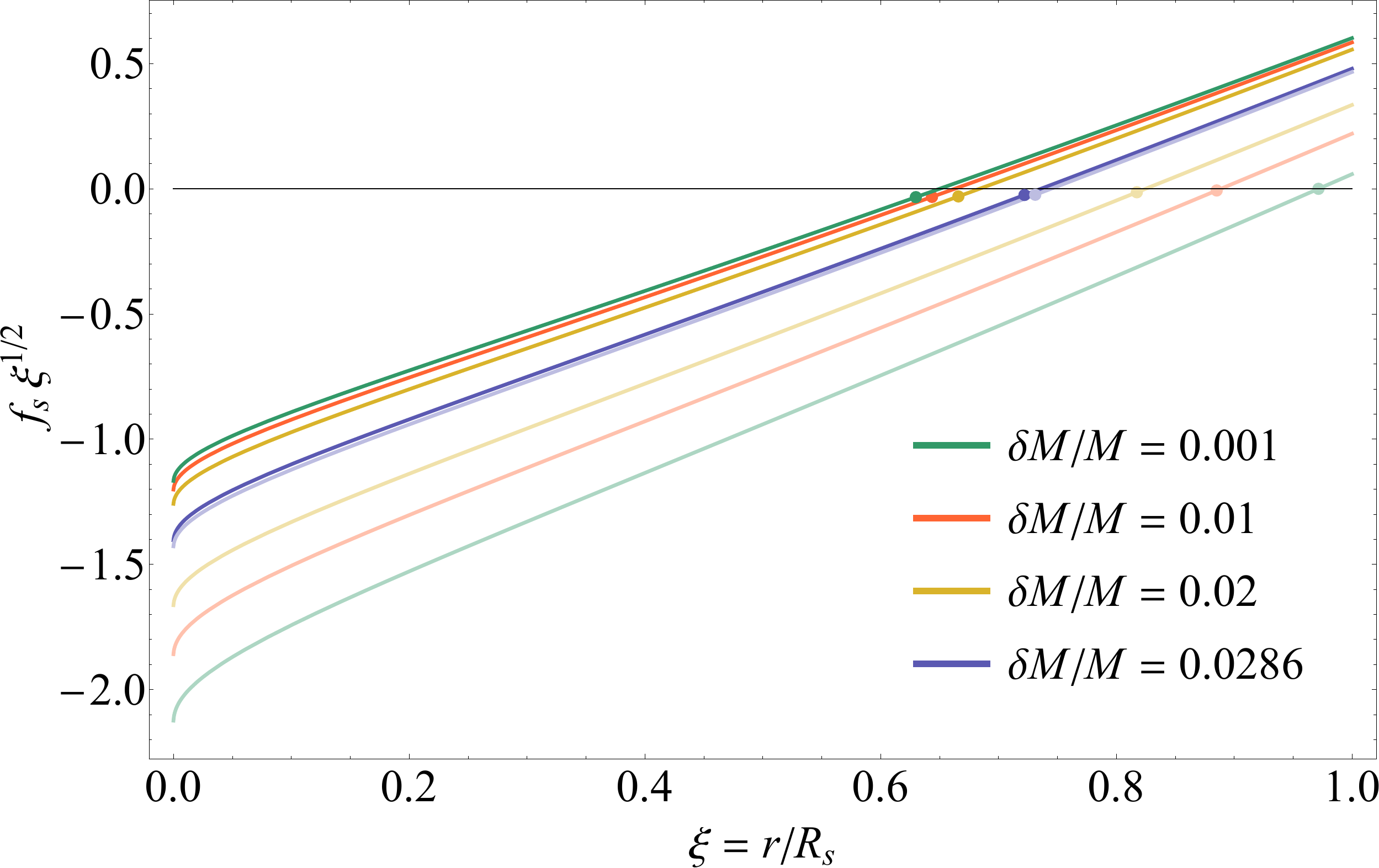} 
   \caption{The self-similar velocity $f_{\rm s}$ normalized by the freefall scaling for both the weak- (light-colored curves) and strong-shock (dark-colored curves) solutions, where here we set $n = 2.5$, $\gamma = 1.4$, and the value of $\delta M/M$ appropriate to each curve is given in the legend. As $\delta M/M$ increases, the weak and strong solutions approach one another, and are nearly equal at $\delta M/M = 0.0286$; above approximately this critical mass loss there are no self-similar solutions for the post-shock flow.}
   \label{fig:fss_dM}
\end{figure}

\begin{figure*}[h!] 
   \centering
   \includegraphics[width=0.495\textwidth]{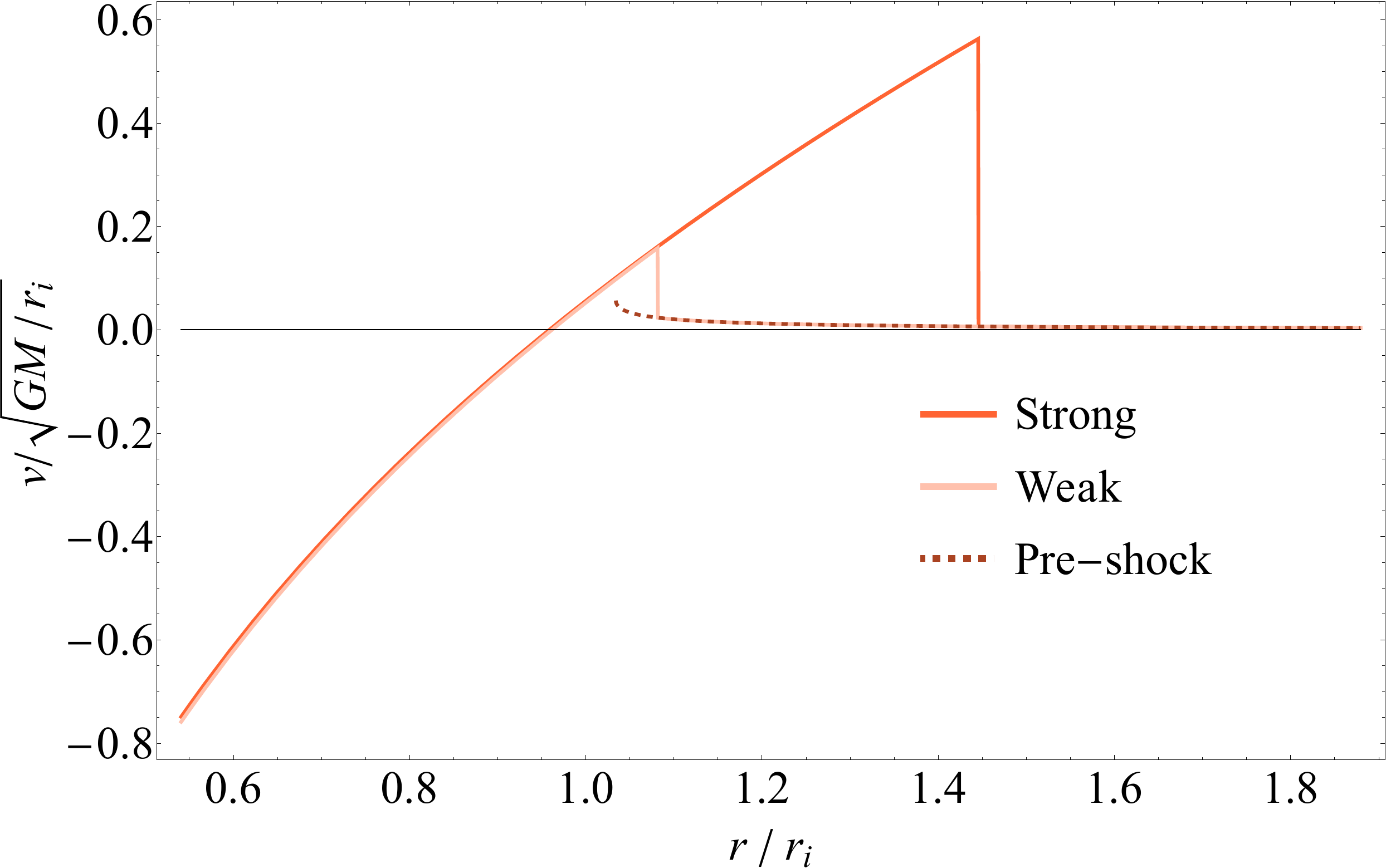} 
 \includegraphics[width=0.495\textwidth]{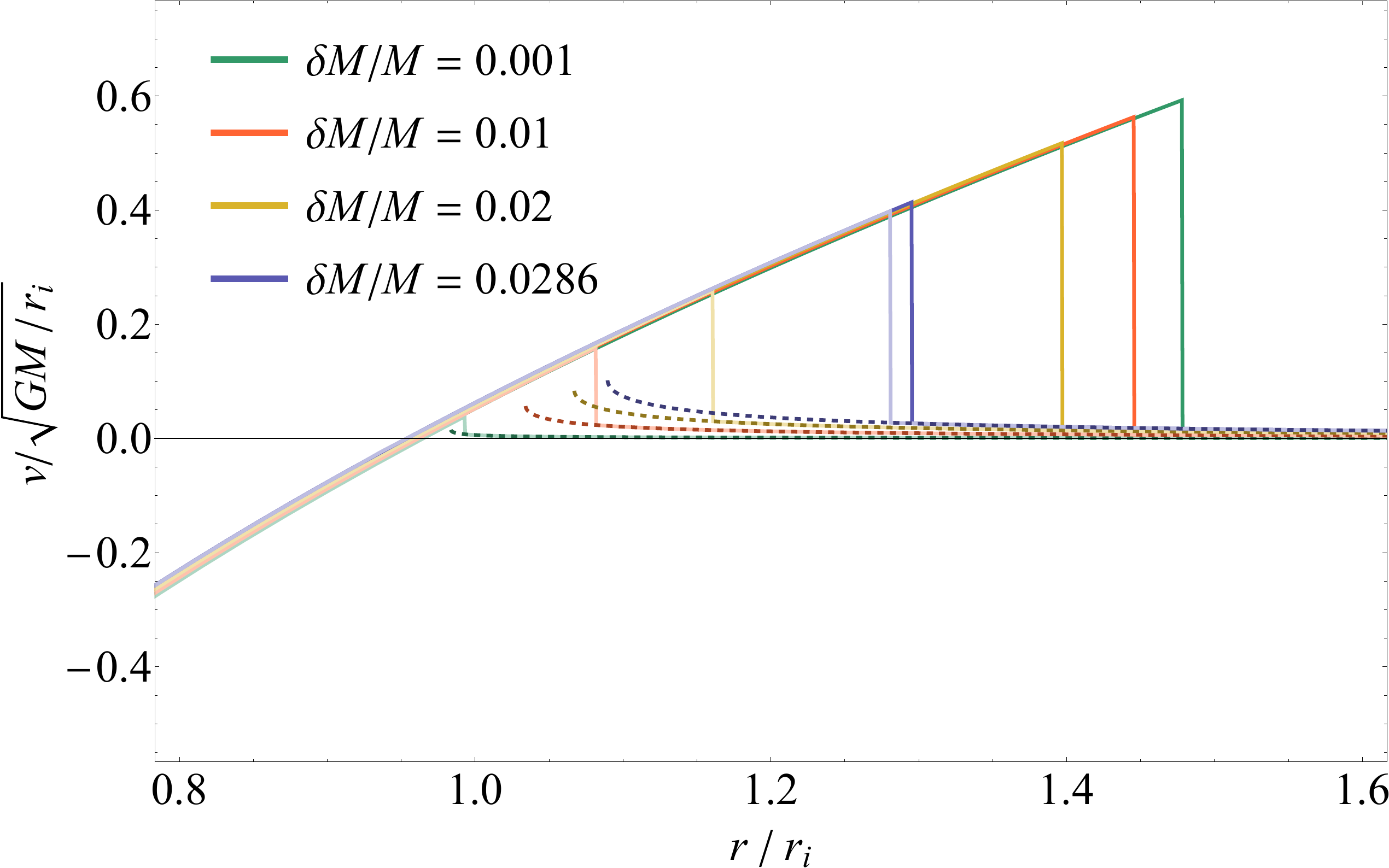} 
   \caption{Left: The velocity normalized by $\sqrt{GM/r_{\rm i}}$ as a function of radius normalized by $r_{\rm i}$ for $n = 2.5$, $\gamma = 1.4$, and $\delta M/M = 0.01$ at a time of $\tau_{\rm i} = r_{\rm i}^{3/2}/\sqrt{GM}$. The dark- and light-colored curves show the strong- and weak-shock solution, respectively, while the dashed curve gives the solution for the pre-shock medium from Section \ref{sec:ambient}. Right: Same as for the left panel, but for the different mass losses given in the legend.}
   \label{fig:strong_weak}
\end{figure*}

\begin{figure*}[h!] 
   \centering
   \includegraphics[width=0.495\textwidth]{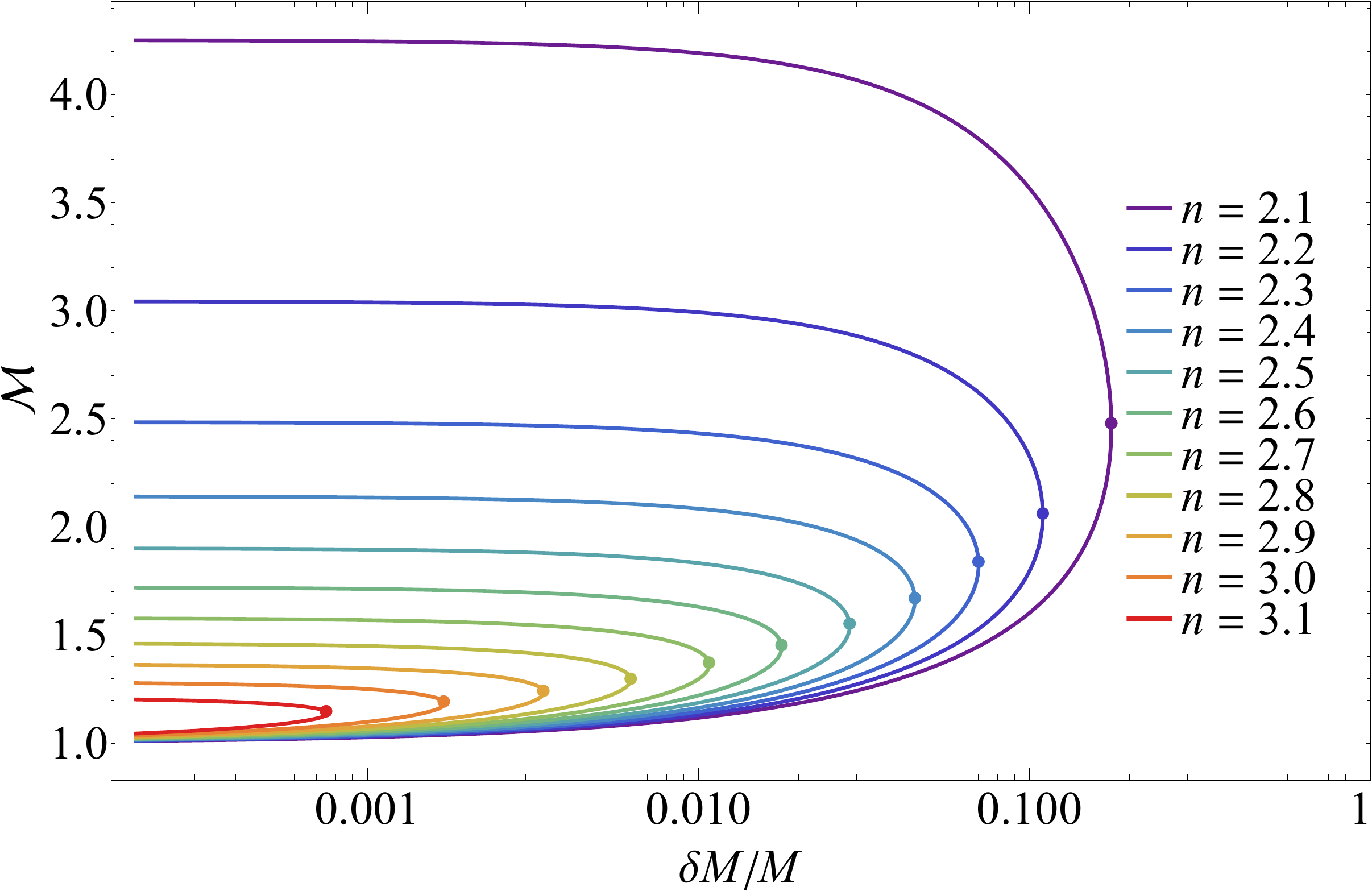} 
    \includegraphics[width=0.495\textwidth]{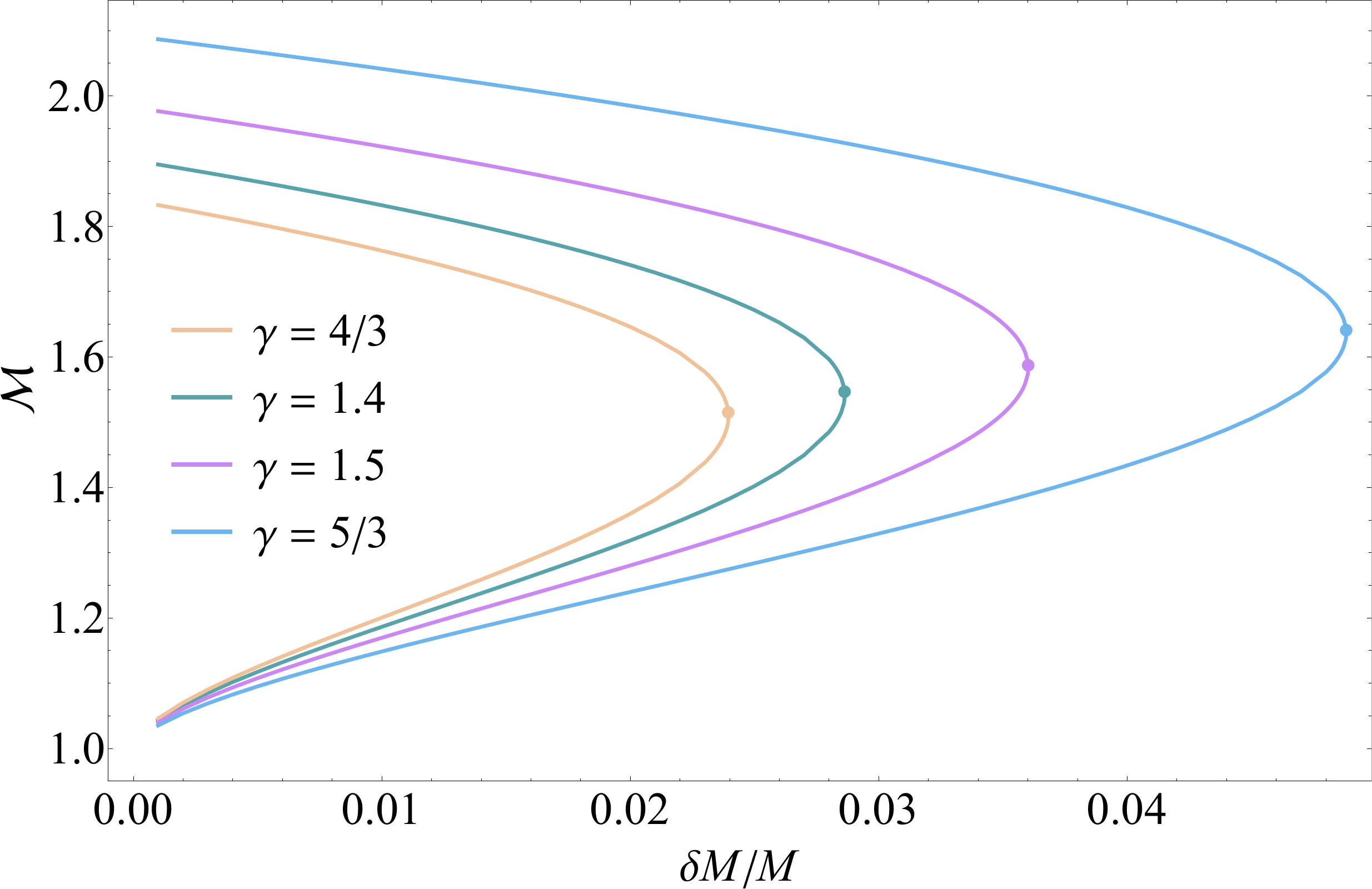} 
   \caption{Left: The shock Mach number as a function of $\delta M/M$ for the power-law indices shown in the legend. For each $\delta M/M$ below a critical value there are two values of the Mach number, one corresponding to the strong-shock solution, the other to the weak-shock solution. As the mass loss increases the two Mach numbers approach one another until they merge to a unique solution at a critical mass loss, that mass loss (and the Mach number at which it occurs) delimited by the points. Above this critical relative mass loss there is no self-similar solution. Right: Same as the left panel, but with $n = 2.5$ and the values of the adiabatic index given in the legend. As $\gamma$ increases, the larger value of the Mach number (for a given $\delta M/M$) increases and the smaller Mach number decreases, and the maximum-possible mass loss (shown by the points) increases.}
   \label{fig:Mach_dM}
\end{figure*}

Figure \ref{fig:fss_dM} shows the self-similar velocity, normalized by the freefall scaling, for $n = 2.5$, $\gamma = 1.4$, and a range of fractional mass losses. As for Figure \ref{fig:fs_gs}, dark curves give the strong-shock solutions, while lightly colored curves correspond to the weak-shock solutions. When the mass loss is small, the weak shock solution approaches the rarefaction-wave limit, in which the velocity is continuous (and equal to zero) at the location of the shock, while the strong-shock solution approaches that described in \citet{coughlin18b}. As $\delta M/M$ increases, the Mach number of the weak-shock solution increases, while that of the strong-shock solution decreases, and at a fractional mass loss of $\delta M/M = 0.0286$ the two solutions are nearly identical. Above $\delta M/M \simeq 0.02863$, we find that there is no self-similar, post-shock solution that matches onto the expanding medium. 

The left panel of Figure \ref{fig:strong_weak} shows the velocity normalized by $\sqrt{GM/r_{\rm i}}$ as a function of $r/r_{\rm i}$ at a time of $t  = \tau_{\rm i} = r_{\rm i}^{3/2}/\sqrt{GM}$ for $n = 2.5$, $\gamma = 1.4$, and $\delta M/M = 0.01$, where the dark, light, and dashed curves give the strong-shock, weak-shock, and pre-shock solutions, respectively. The right panel shows the same solutions but for the values of the mass loss shown in the legend, and highlights the fact that as the mass loss approaches the critical value of $\delta M/M \simeq 0.02863$, the weak- and strong-shock solutions approach one another. At a given time, the strong-shock solution exceeds the weak-shock solution owing to its larger Mach number.

The left panel of Figure \ref{fig:Mach_dM} illustrates the Mach number as a function of $\delta M/M$ for the power-law indices in the legend, and we set $\gamma = 1+1/n$. Below a critical $\delta M/M$ there are two Mach numbers, one (the larger) corresponding to the strong-shock solution, the other (smaller) corresponding to the weak-shock solution. As $\delta M/M$ increases the strong- and weak-shock Mach numbers converge toward one another, until they equal one another at a single value for a critical relative mass loss, that mass loss (and Mach number) shown by the point on each curve. Above this mass loss there are no self-similar solutions for the shock. The right panel of Figure \ref{fig:Mach_dM} shows the shock Mach number for $n = 2.5$ and the $\gamma$ shown in the legend. As the adiabatic index increases, the maximum Mach number increases and the minimum Mach number decreases, and the maximum mass loss (above which there are no self-similar solutions) is pushed to larger values. 

\begin{figure*}[t!] 
   \centering
   \includegraphics[width=0.505\textwidth]{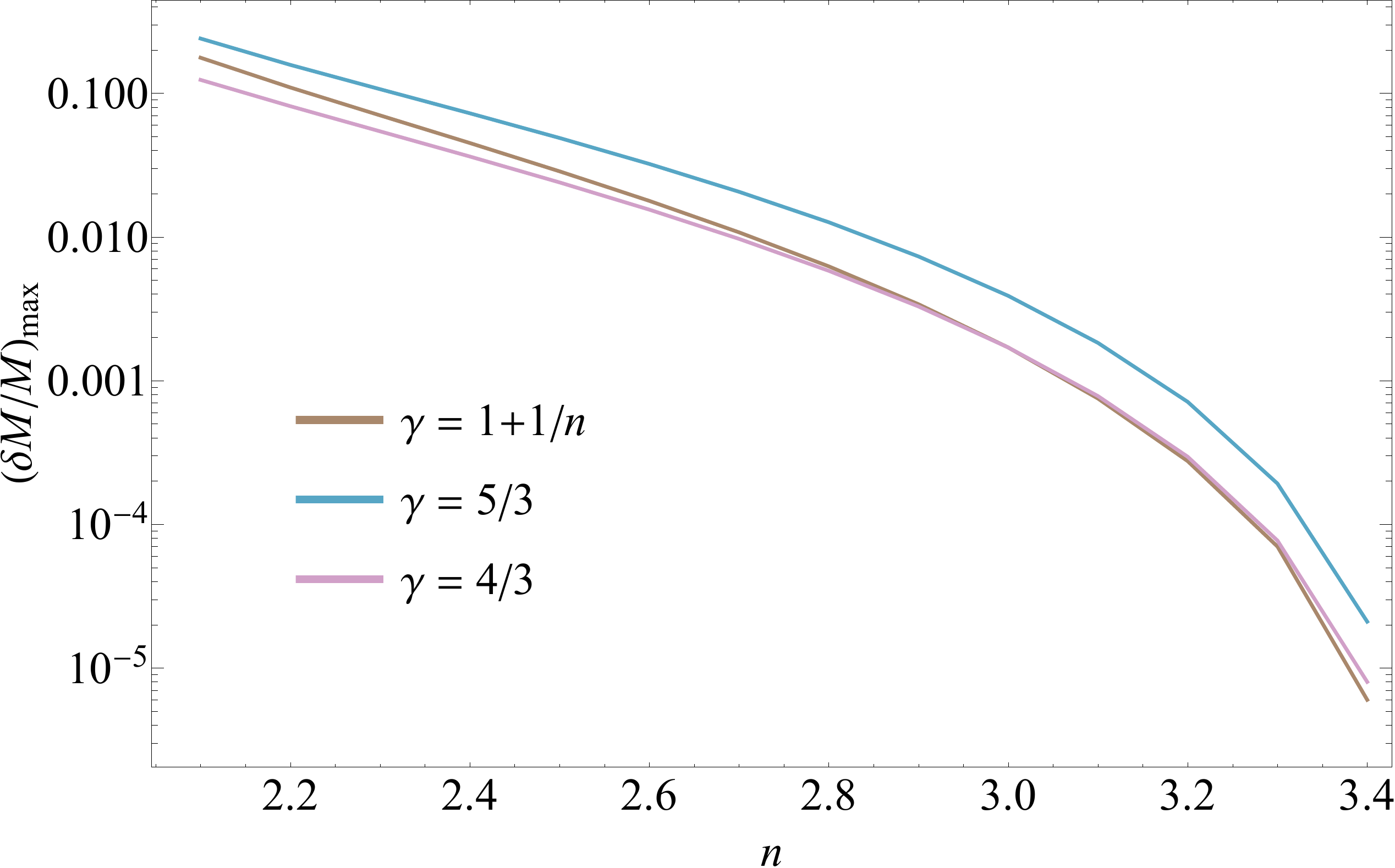} 
    \includegraphics[width=0.475\textwidth]{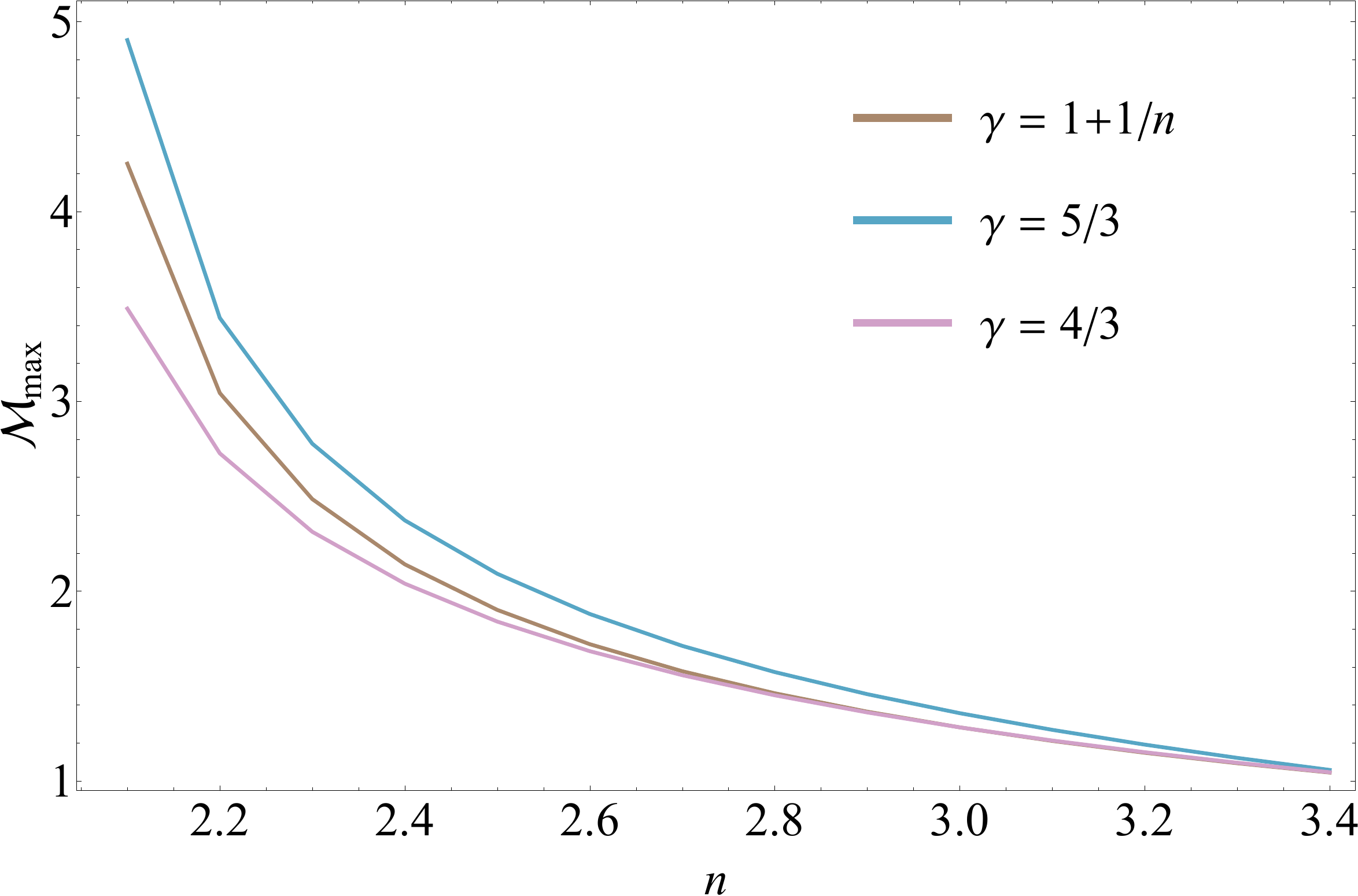}
   \caption{The maximum mass loss (left) and the maximum Mach number (right) -- above which there are no self-similar solutions for the post-shock gas -- as a function of the power-law index of the ambient medium and the adiabatic indices given in the legend. }
   \label{fig:dM_max}
\end{figure*}

The left panel of Figure \ref{fig:dM_max} shows the maximum relative mass loss -- above which there are no self-similar solutions -- as a function of the power-law index of the ambient medium and the adiabatic indices shown in the legend. As $n$ approaches 3.5, the maximum mass loss approaches smaller values, and there are no solutions (with finite $\delta M/M$) for $n \ge 3.5$. The right panel of this figure shows the maximum Mach number above which there are no solutions, which corresponds to a fractional mass loss of zero (see the left panel of Figure \ref{fig:Mach_dM}). As $n$ increases the Mach number declines, approaching and equaling unity for $n = 3.5$. As $n$ decreases the maximum Mach number increases, and grows unbounded in the limit that $n \rightarrow 2$. Table \ref{tab:Machmax} gives the values of the maximum Mach number and the maximum fractional mass loss for different power-law and adiabatic indices.

As for the dynamic and self-similar pre-shock solutions (Sections \ref{sec:basic} and \ref{sec:ambient}, respectively), we can calculate the kinetic energy contained in the expanding, post-shock gas. While the lower integration limit could extend to zero, in general we expect any explosion produced by the mass loss to contain predominantly the outward-moving gas at any given time. Gas with positive velocity is confined by the shock front (at $\xi = 0$) and the self-similar radius where the velocity equals zero, which we define to be $\xi_0$, i.e., $f_{\rm s}(\xi_0) = 0$. Then the kinetic energy of this gas is
\begin{equation}
\begin{split}
E_{\rm kin} &= 4\pi\int_{R_0(t)}^{R_{\rm sh}(t)}\frac{1}{2}v^2\rho r^2 dr \\
&=E_{\rm i}\tau^{\frac{2}{3}\left(2-n\right)}\frac{2}{3}\eta_{\rm sh}^{2\left(n-5\right)/3}\int_{\xi_0}^{1}f_{\rm s}^2g_{\rm s}\xi^2d\xi,
\end{split}
\end{equation}
where, as for the previous section,
\begin{equation}
E_{\rm i} = \frac{4\pi}{3}GM\rho_{\rm i}r_{\rm i}^2, \quad \tau = \frac{\sqrt{GM}t}{r_{\rm i}^{3/2}}.
\end{equation}
In a failed supernova we expect the scale radius of the envelope, $r_{\rm i}$, to coincide with the minimum location of the shock (see Section \ref{sec:discussion} for additional discussion), and hence the maximum kinetic energy contained in the post-shock gas occurs when $R_{\rm sh} = r_{\rm i}$, or when $\tau = \eta_{\rm sh}$, and hence 
\begin{equation}
E_{\rm kin, max} = \frac{2}{3}E_{\rm i}\eta_{\rm sh}^{-2}\int_{\xi_0}^{1}f_{\rm s}^2g_{\rm s}\xi^2d\xi. \label{Ekmax}
\end{equation}

\begin{figure}[htbp] 
   \centering
   \includegraphics[width=0.47\textwidth]{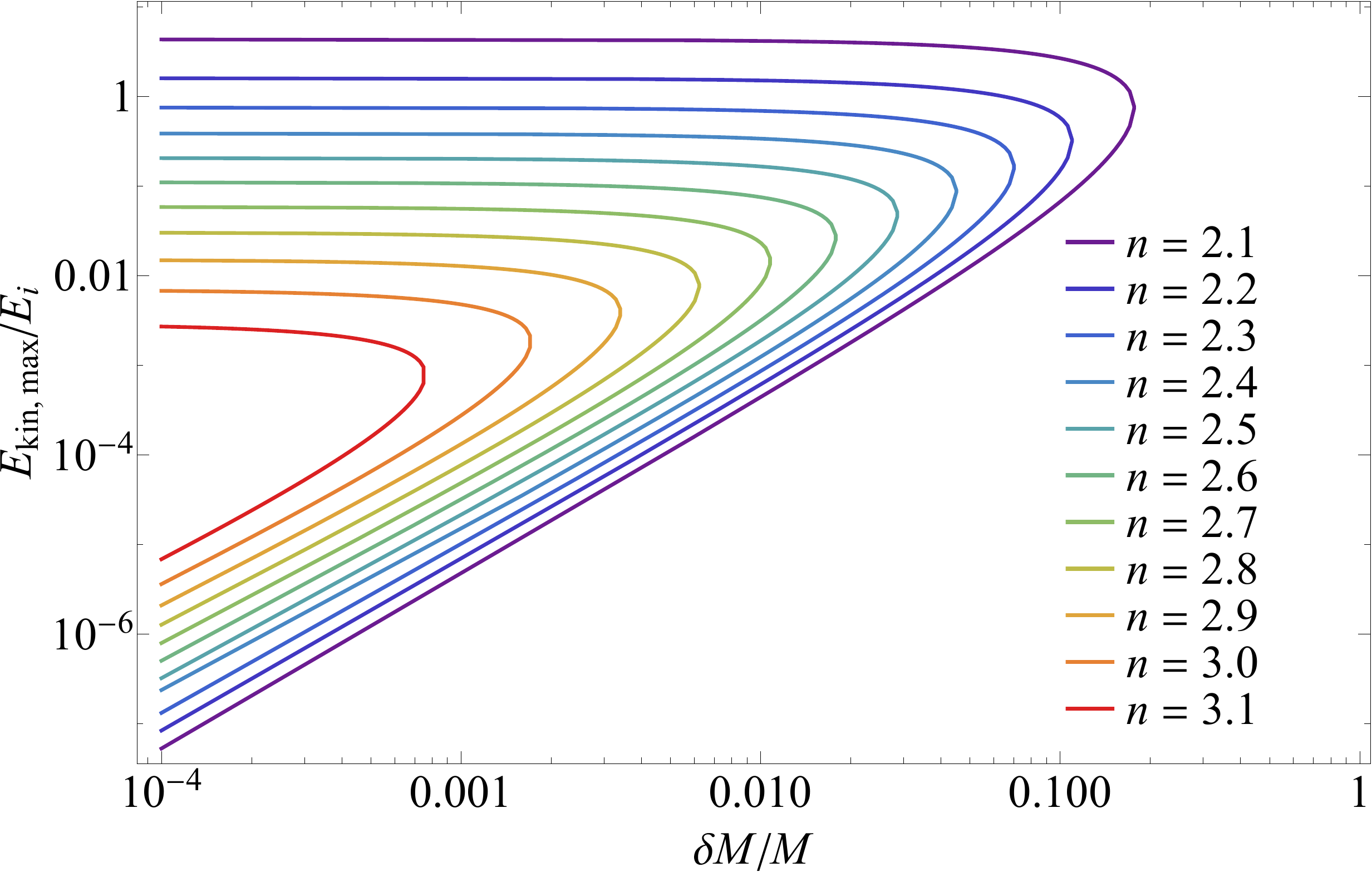} 
   \caption{The maximum kinetic energy contained in the outward-moving gas in the post-shock flow as a function of the fractional mass loss. Each curve corresponds to the power-law index of the ambient medium shown in the legend, and here we set $\gamma = 1+1/n$. For each mass loss less than the maximum-possible value, which coincides with the right-most point on each curve (see also Figure \ref{fig:dM_max} and Table \ref{tab:Machmax}), there are two values for the kinetic energy, the smaller (larger) of which corresponds to the weak (strong) solution. }
   \label{fig:Ekmax_shock}
\end{figure}

Figure \ref{fig:Ekmax_shock} shows the maximum kinetic energy contained in the post-shock gas as a function of the fractional mass loss, where the power-law index of the ambient medium appropriate to each curve is given in the legend and we set $\gamma = 1+1/n$. For each $\delta M/M$ below the maximum (i.e., the right-most point on each curve) there are two values of the kinetic energy, the smaller (larger) of which corresponds to the weak (strong) solution. As the value of $n$ increases, the maximum kinetic energy contained in the strong branch declines, while that in the weak branch increases, until the two merge at the point that coincides with the maximum-possible fractional mass loss. For values appropriate to the YSG in \citet{fernandez18}, we find $E_{\rm kin, max} = \{3.2\times10^{46}, \, 3.6\times 10^{44}\}$ erg for the strong-shock and weak-shock kinetic energies, respectively; the former is in good agreement with that recovered numerically by \citet{fernandez18}, being $4\times 10^{46}$ erg. 

\begin{table*}[t!]
   \begin{tabular}{|c|c|c|c|} 
   \hline
   \{$\mathscr{M}_{\rm max},(\delta M/M)_{\rm max}$\} & $\gamma = 1+1/n$ & $\gamma = 5/3$ & $\gamma = 4/3$ \\
   \hline
   $n = 2.1$ & \{4.25, 0.177\} & \{4.90, 0.240\} & \{3.49, 0.124\} \\
   \hline
    $n = 2.2$ & \{3.04, 0.110\} & \{3.44, 0.158\} & \{2.73, 0.0813\} \\
   \hline
    $n = 2.3$ & \{2.48, 0.0701\} & \{2.78, 0.107\} & \{2.31, 0.0543\} \\
   \hline
    $n = 2.4$ & \{2.14, 0.0450\} & \{2.37, 0.0725\} & \{2.04, 0.0363\} \\
   \hline
    $n = 2.5$ & \{1.90, 0.0286\} & \{2.09, 0.0488\} & \{1.84, 0.0240\} \\
   \hline
    $n = 2.6$ & \{1.72, 0.0178\} & \{1.88, 0.0322\} & \{1.68, 0.0155\} \\
   \hline
    $n = 2.7$ & \{1.58, 0.0108\} & \{1.71, 0.0206\} & \{1.57, 0.00969\} \\
   \hline
    $n = 2.8$ & \{1.46, 0.00624\} & \{1.57, 0.0127\} & \{1.45, 0.00581\} \\
   \hline
    $n = 2.9$ & \{1.36, 0.00340\} & \{1.46, 0.00731\} & \{1.36, 0.00328\} \\
   \hline
    $n = 3.0$ & \{1.28, 0.0017\} & \{1.36, 0.00389\} & \{1.28, 0.0017\} \\
   \hline
    $n = 3.1$ & \{1.21, 0.00075\} & \{1.27, 0.00183\} & \{1.21, 0.00078\} \\
   \hline
    $n = 3.2$ & \{1.15, 0.000275\} & \{1.19, 0.000711\} & \{1.15, 0.000295\} \\
   \hline
    $n = 3.3$ & \{1.09, $7.0\times10^{-5}$\} & \{1.12, 0.000192\} & \{1.10, $7.7\times 10^{-5}$\} \\
   \hline
    $n = 3.4$ & \{1.04, $6.0\times 10^{-6}$\} & \{1.06, $2.1\times10^{-4}$\} & \{1.05, $8.0\times 10^{-6}$\} \\
   \hline
   \end{tabular}
   \caption{The maximum Mach number $\mathscr{M}_{\rm max}$ and mass loss $\left(\delta M/M\right)_{\rm max}$ above which no self-similar solution exists. The power-law index of the ambient medium is given in the left column, while the adiabatic index is shown in the top row.}
   \label{tab:Machmax}
\end{table*}

\section{Discussion and Interpretation}
\label{sec:discussion}
In this section we discuss the features of the previously described solutions and the implications for failed supernovae and their fallback accretion.

\subsection{Stability}
\label{sec:stability}
The stronger-shock solutions described in the previous section represent generalizations of the self-similar solutions found in \citet{coughlin18b}. \citet{coughlin19c} showed that the \citet{coughlin18b} solutions are very weakly unstable, with instabilities growing as power-laws in time and power-law indices $\lesssim 0.2$. Since the solutions found here approach the \citet{coughlin18b} solutions in the limit that $\delta M/M \rightarrow 0$, we conclude that the stronger-shock solutions are also unstable to radial perturbations (in the form of, e.g., small changes in the Mach number of the shock). Similarly, our weak-shock solutions are generalizations to the rarefaction-wave solutions described in \citet{coughlin19c}, to which our solutions tend in the limit that $\delta M/M \rightarrow 0$. Since the rarefaction-wave solutions are stable \citep{coughlin19c}, we conclude that the weak-shock solutions are also stable. 

Our interpretation of this stability vs.~instability is that, if the shock has a Mach number greater than that of the strong-shock solution, $\mathscr{M}_{\rm sh, s}$, owing to, e.g., initial conditions, it will continue to strengthen (i.e., the shock Mach number will grow) and approach the Sedov-Taylor blastwave with a conserved energy. On the other hand, if the shock Mach number is less than $\mathscr{M}_{\rm sh, s}$ and greater than $\mathscr{M}_{\rm sh, w}$ at any point, it will weaken and asymptotically approach the Mach number of the weak-shock solution, $\mathscr{M}_{\rm sh, w}$. In this way, $\mathscr{M}_{\rm sh, s}$ represents the critical Mach number necessary to generate a strong (i.e., energy-conserving) explosion.

The growth rate of the perturbations can be derived in a way that parallels the approach in \citet{coughlin19c}: we write the shock position as a function of time implicitly as
\begin{equation}
\frac{\sqrt{GM}t}{R_{\rm sh}^{3/2}} = \eta_{\rm sh}+\eta_1(\tau), \label{Rshpert}
\end{equation}
where $\tau = \ln(R_{\rm sh})$. We also write $v = V_{\rm sh}\left\{f_0(\xi)+f_1(\xi,\tau)\right\}$ and analogously for the self-similar density and pressure, assume all subscript-1 quantities are small enough such that their products can be ignored, and linearize the fluid equations and the boundary conditions. The first-order equations and boundary conditions can be satisfied if the time dependence of the first-order quantities is written as $e^{\sigma\tau}$, i.e., $\eta(\tau) \propto e^{\sigma\tau}$, $f_1(\xi,\tau) = e^{\sigma\tau}f_1(\xi)$, etc., and the eigenvalue $\sigma$ is constrained by requiring that the first-order solutions smoothly pass through the sonic point.

While there will be an infinite number of eigenvalues, one will be close to zero and another will be close to $-3/2$. The reason for this is that, if there were neither a velocity nor spatial scale, meaning that the position and velocity of the shock at $t = 0$ were arbitrary, then the constraint on the self-similarity of the solution would be only related to the acceleration of the shock via (e.g., \citealt{coughlin22})
\begin{equation}
\frac{R\dot{V}}{V^2} = -\frac{1}{2} \,\,\, \Rightarrow \,\,\,  R = \left(C_0t+C_1\right)^{2/3},
\end{equation}
where $C_0$ and $C_1$ are constants of integration related to the initial shock velocity and position, respectively. If the self-similar solution is defined relative to a specific value of $C_0$ and $C_1$, then we can let $C_0 \rightarrow C_0+\delta C_0$ and $C_1 \rightarrow C_1+\delta C_1$, and Taylor expanding the above solution to leading order in $\delta C_0$ and $\delta C_1$ shows that
\begin{equation}
\begin{split}
R &= R_0(t)\left(1 + \delta C_0 + \delta C_1 R_0^{-3/2}\right) \\ 
&= R_0(t)\left(1+\delta C_0e^{\sigma_0\tau}+\delta C_1e^{\sigma_1\tau}\right).
\end{split}
\end{equation}
We thus see that there are two ``trivial'' eigenvalues, given by $\sigma_0 = 0$ and $\sigma_1 = -3/2$, that correspond to renormalizations of the shock velocity and position. Since neither the shock velocity nor position is arbitrary in the context of the self-similar solutions presented here -- the escape speed provides a fundamental velocity scale and the position of the shock at $t = 0$ must be zero for the self-similar solution -- the two eigenvalues will not be exactly zero and $-3/2$, but we expect them to be close to these values, especially because they are exactly $-3/2$ and $\lesssim 0.2$ when the mass loss is zero \citep{coughlin19c}.

As the mass loss increases and approaches $\left(\delta M/M\right)_{\rm max}$, $\mathscr{M}_{\rm sh, s}$ decreases, implying that a smaller Mach number is required for the shock to transition into the energy-conserving regime. As $\delta M/M \rightarrow \left(\delta M/M\right)_{\rm max}$, the weak-shock solution must become unstable, and the fact that the strong-shock and weak-shock solutions converge implies that the eigenvalue describing the instability must be a repeated root at $\sigma = 0$. Thus, perturbations to the solution with $\delta M/M = \left(\delta M/M\right)_{\rm max}$ grow logarithmically with the shock radius (and with time). 

For fractional mass losses above $\left(\delta M/M\right)_{\rm max}$, there are no solutions that satisfy $R_{\rm sh} \propto t^{2/3}$ and that yield accretion onto the black hole. This implies that for mass losses above this value, the shock must transition to the energy-conserving/Sedov-Taylor regime. Given the expected, extremely slow growth of the perturbations to the self-similar solutions, this transition likely takes place over many orders of magnitude in the initial radius of the shock, and could be modeled analytically by using the $\left(\delta M/M\right)_{\rm max}$ self-similar solutions as the background solution and letting the perturbations be driven by the excess mass loss for which the self-similar solutions cannot account.

Finally, the stability we have discussed so far concerns only radial perturbations, but there is also the possibility of angular perturbations. Accounting for these amounts to writing the corrections to the fluid quantities as spherical harmonics and also accounting for the angular components of the velocity (e.g., \citealt{ryu87}). In a subsequent paper we plan to rigorously analyze the stability of the self-similar solutions and make comparisons to hydrodynamical simulations that explore the transition to the energy-conserving regime.

\subsection{Solutions with $n \le 2$ and $n \ge 3.5$}
\label{sec:nle2}
As $n$ approaches 2, the maximum Mach number and the maximum mass loss both increase (see Figure \ref{fig:Mach_dM}) and diverge in the limit that $n \rightarrow 2$, the reason being that the binding energy of the envelope diverges in this limit and is infinite for all $n \le 2$. Thus, for $n \le 2$, the strong explosion/Sedov-Taylor self-similar solution is ``unstable'' to the presence of a gravitational field and will inevitably decelerate to the point that gravity is not ignorable. Said another way, the velocity for an energy-conserving shock varies as $V \propto R^{(n-3)/2}$, and $(n-3)/2 < -1/2$ for all $n < 2$. Therefore, an energy-conserving shock will always reach a point where the shock speed is comparable to the freefall speed if $n < 2$. 

\begin{figure}[htbp] 
   \centering
   \includegraphics[width=0.475\textwidth]{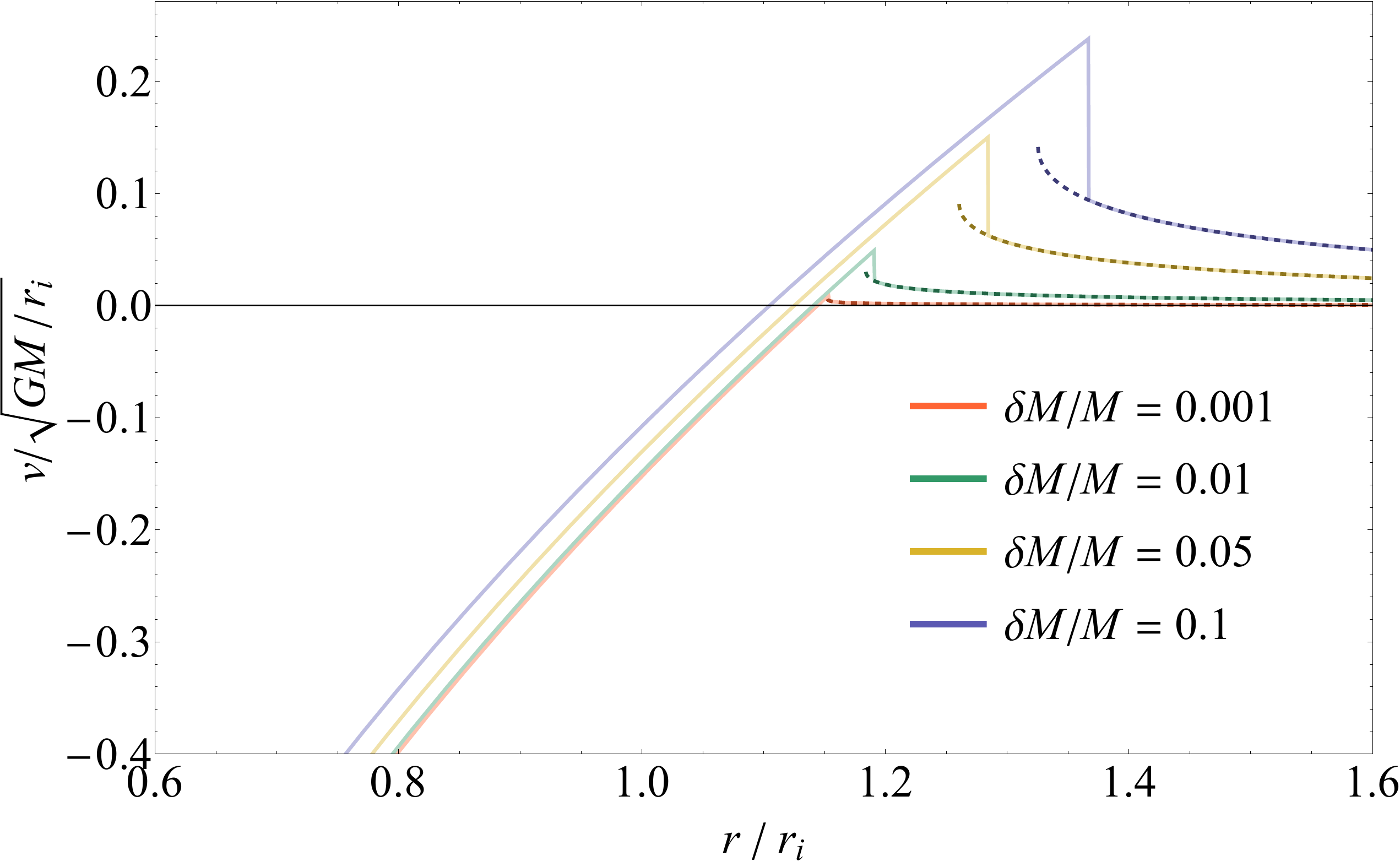} 
   \caption{The velocity as a function of radius for the weak-shock solutions when $n = 1.5$, $\gamma = 5/3$, and the mass loss is given by the values in the legend. Because $n < 2$, the stronger-shock (and unstable) solution does not exist, whereas the weak-shock solution (shown by the lightly colored curves here; the ambient velocity profile is shown by each dashed curve) exists even for very large fractional mass losses. }
   \label{fig:nle2}
\end{figure}

As $n \rightarrow 2$, the critical Mach number $\mathscr{M}_{\rm sh, s}$ that delimits the transition to the energy-conserving regime approaches infinity, and the strong-shock solution does not exist for $n < 2$. However, \citet{coughlin19c} noted that the rarefaction-wave solution still exists in this regime of parameter space, and we expect the same to be true here -- that the weak shock solution is the only stable solution in this limit. We find that this is indeed the case: Figure \ref{fig:nle2} shows the velocity as a function of radius for the weak-shock solutions with $n = 1.5$, $\gamma = 5/3$ ($=1+1/n$), and the fractional mass losses in the legend. We find that there is no stronger-shock solution, while the weak-shock solution exists -- even when the fractional mass loss is set to the large value of 0.1. When the mass loss is small, the weak-shock solution approaches the rarefaction-wave limit. 

On the other hand, as the power-law index of the ambient medium increases, $\mathscr{M}_{\rm sh, s}$ decreases, implying that it is easier for the shock to transition to the strong regime; this is consistent with the fact that the binding energy of the ambient medium (from any finite $r_{\rm i}$) declines as the power-law index grows. In the limit that $n \rightarrow 3.5$, the strong-shock and weak-shock Mach numbers approach one another, and equal 1 at $n = 3.5$. The maximum fractional mass loss above which there are only strong and energy-conserving solutions also approaches zero at $n = 3.5$. 

These results imply that any finite value of the mass loss, no matter how small, will result in the strengthening of the shock into the energy-conserving regime if the density profile falls off more steeply than $\propto r^{-3.5}$. The physical importance of the power-law index of 3.5 is not clear, but as shown in \citet{coughlin19c}, the rarefaction-wave solution is unstable in this case.

\subsection{Implications for failed supernovae}
\label{sec:implications}
During the core collapse of a massive star, the overlying envelope is informed of the loss of pressure support by the propagation of a rarefaction wave. Thus, at the time the mass loss occurs, the rarefaction wave will be at some radius within the star $r_{\rm i}$. If that radius coincides with a location within the star at which the dynamical time is greater than the timescale over which the neutrinos reduce the mass of the core, then it is reasonable to assume that the mass loss occurs effectively instantaneously, and -- provided that the density profile can be well-approximated by a power-law -- the surrounding envelope will expand in the reduced gravitational field in a way that mimics the solutions presented in Section \ref{sec:ambient}.

\begin{figure}[htbp] 
   \centering
   \includegraphics[width=0.47\textwidth]{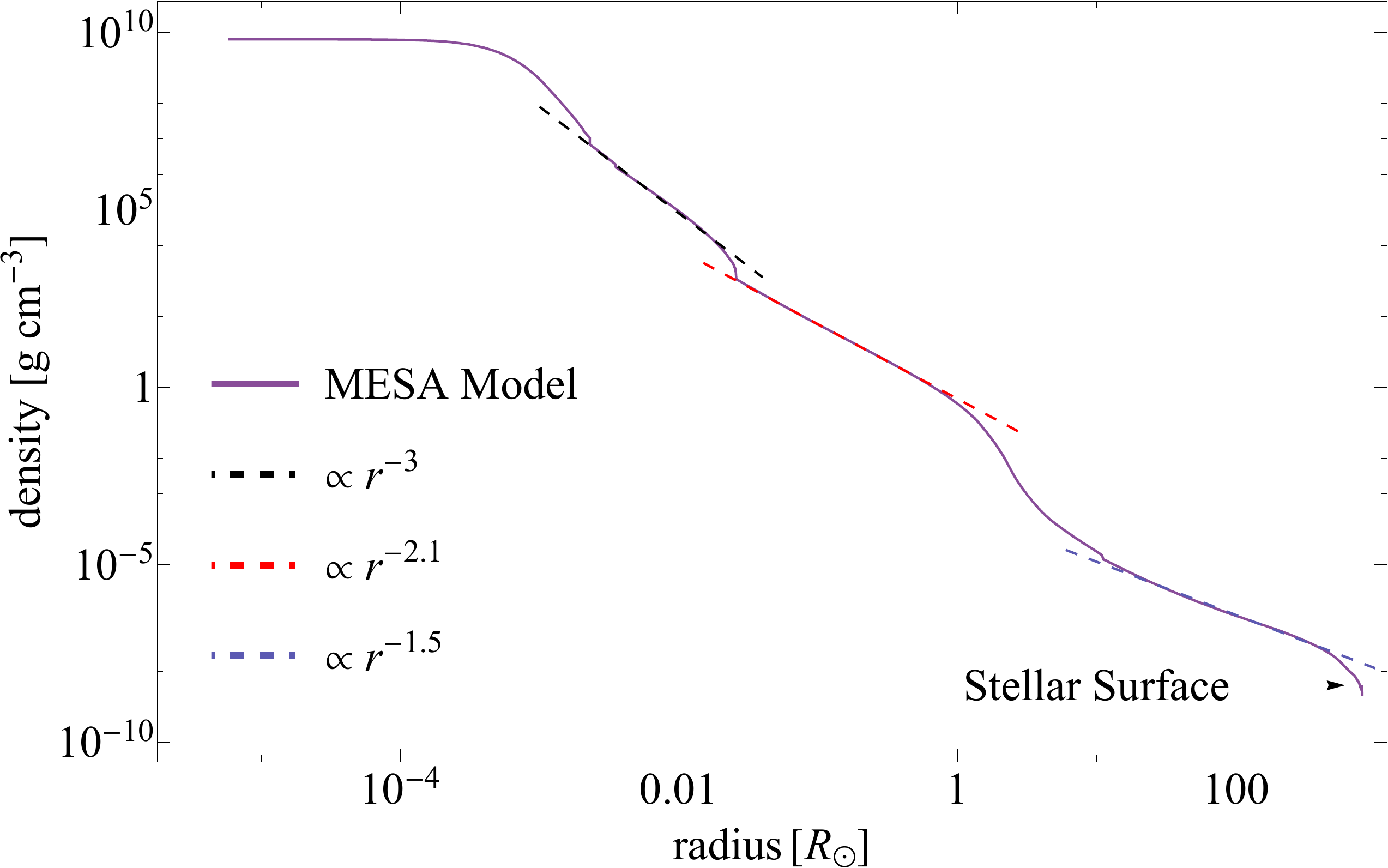} 
   \caption{The density profile of a 20 $M_{\odot}$ star evolved to the onset of core collapse with the stellar evolution code {\sc mesa}. The various dashed lines illustrate regions in the star that are well-approximated by power-laws.}
   \label{fig:mesa_star}
\end{figure}

We then expect the rarefaction wave to transition into a shock that joins the expanding envelope to the accreting gas. Because an evolved star consists of (at the simplest level) nested nuclear-burning shells, it may be that when the shock initially forms, it is in a region of the star that has a very steep density profile. For example, Figure \ref{fig:mesa_star} shows a $20 M_{\odot}$ star evolved to core collapse with the stellar evolution code {\sc mesa} \citep{paxton11, paxton13}, and outside of the core the density profile can be well-modeled by piecewise-continuous power-laws. If the rarefaction wave is initially in the region that is characterized by $\rho \propto r^{-3}$, then it is likely that the fractional mass loss will be larger than $\left(\delta M/M\right)_{\rm max} \simeq 0.0017$ (assuming $\gamma = 4/3$; see Table \ref{tab:Machmax}). 

If this is the case, then the shock Mach number will continue to grow and the shock will strengthen into the strong and energy-conserving regime while it is in this steep density profile. However, once the shock reaches the radius at which the density is $\propto r^{-2.1}$, the critical mass loss $\left(\delta M/M\right)_{\rm max} \simeq 0.177$ (again, Table \ref{tab:Machmax}) is likely too large to be achieved through the neutrino mass loss mechanism, and hence the ability of the shock to remain strong (and continue to strengthen) depends on whether the shock Mach number is initially greater than $\mathscr{M}_{\rm sh, s}$ for that value of $n$. From Figure \ref{fig:Mach_dM}, this Mach number is $\sim 4$. If the Mach number exceeds this critical value, then the Mach number will continue to increase, while if it is smaller than $\mathscr{M}_{\rm sh, s}$, the Mach number will decline and approach $\mathscr{M}_{\rm sh, w}$. 

For the density profile in Figure \ref{fig:mesa_star}, the hydrogen envelope has a relatively shallow power-law index and satisfies $\rho \propto r^{-1.5}$. Since this is less steep than $r^{-2}$, the shock Mach number will decline in this region owing to the non-existence of strong-shock (either with $\mathscr{M}_{\rm sh, s}$ or the Sedov-Taylor blastwave) solutions and asymptotically approach the Mach number of the weak shock appropriate to that power-law and fractional mass loss (see Figure \ref{fig:nle2}). Moreover, because the density power-law index is shallow, the mass contained interior to the shock will increase non-trivially. We therefore expect the shock Mach number to decline gradually owing to the change in the fractional mass loss with radius.

Finally, upon encountering the edge of the stellar envelope where the density profile drops precipitously, the shock will start to accelerate down the steep density gradient. As we saw above, once the power-law index steepens past $\propto r^{-3.5}$, there are only strong (energy-conserving) solutions that characterize the propagation of the shock. In such ultra-steep density profiles the shock enters an accelerating regime when it is strong (Mach number much greater than one; \citealt{waxman93}), and as it reaches the edge it accelerates rapidly \citep{sakurai60}. Here, however, the shock will originally be weak, and it is not clear how rapidly these accelerating regimes are reached, if at all, when this is the case. 

In a realistic core-collapse stellar progenitor, we therefore expect the shock to be characterized by distinct regimes of increasing or decreasing Mach number depending on the steepness of the density profile and the fractional mass loss. If the density profile of the outer, hydrogen envelope is sufficiently steep and the mass loss (and/or initial Mach number) is large enough upon encountering the outer envelope, the neutrino mass loss will result in a shock that strengthens in Mach number and produces a relatively strong explosion. It is likely that one can interpolate between the weak and strong-shock regimes by using the self-similar solutions presented here, which was done in the case of strong and energy-conserving explosions by \citet{matzner99}. If one also incorporates the results of a perturbation analysis to derive the non-self-similar corrections to the shock propagation (see Section \ref{sec:stability}), then one can account self-consistently for the initial shock position and velocity, and also for the time-dependent deviation away from or the convergence toward the strong and weak-shock solutions. We defer such an exercise alongside comparisons to numerical simulations to future work.

\subsection{Fallback accretion}
\label{sec:fallback}
The self-similar solutions produce only bound material to the black hole, implying that each fluid element hit by the shock will eventually fall back to the origin. This conclusion follows from the fact that the self-similar shock position and velocity vary as $R_{\rm sh} \propto t^{2/3}$ and $V_{\rm sh} \propto t^{-1/3}$, and if a fluid element hit by the shock were to become unbound, it would attain a constant velocity at sufficiently large radii and overtake the shock. 

The accretion rate onto the black hole is given by
\begin{equation}
\begin{split}
\dot{M} &= -4\pi \lim_{r\rightarrow 0}\left[r^2\rho v\right] \\
&=\frac{2}{3}\dot{M}_{\rm i}\tau^{1-2n/3}\eta_{\rm sh}^{2n/3-2}\lim_{\xi \rightarrow 0}\left[-\xi^2f_{\rm s}g_{\rm s}\right],
\end{split}
\end{equation}
where
\begin{equation}
\dot{M}_{\rm i} = 4\pi r_{\rm i}^2\rho_{\rm i}\sqrt{\frac{GM}{r_{\rm i}}}.
\end{equation}
Thus, the fallback rate onto the black hole varies with time as a power-law, and the power-law index is shallower than $-5/3$ for all $n < 4$ (also showing that the self-similar solutions do not produce any unbound gas). Since we expect the self-similar solution to hold only when the shock is at radii greater than $r_{\rm i}$, the maximum accretion rate (assuming $n > 3/2)$ onto the black hole is
\begin{equation}
\dot{M}_{\rm max} = \dot{M}_{\rm i}\times \frac{2}{3\eta_{\rm sh}}\lim_{\xi \rightarrow 0}\left[-\xi^2f_{\rm s}g_{\rm s}\right].
\end{equation}

\begin{figure}[htbp] 
   \centering
   \includegraphics[width=0.475\textwidth]{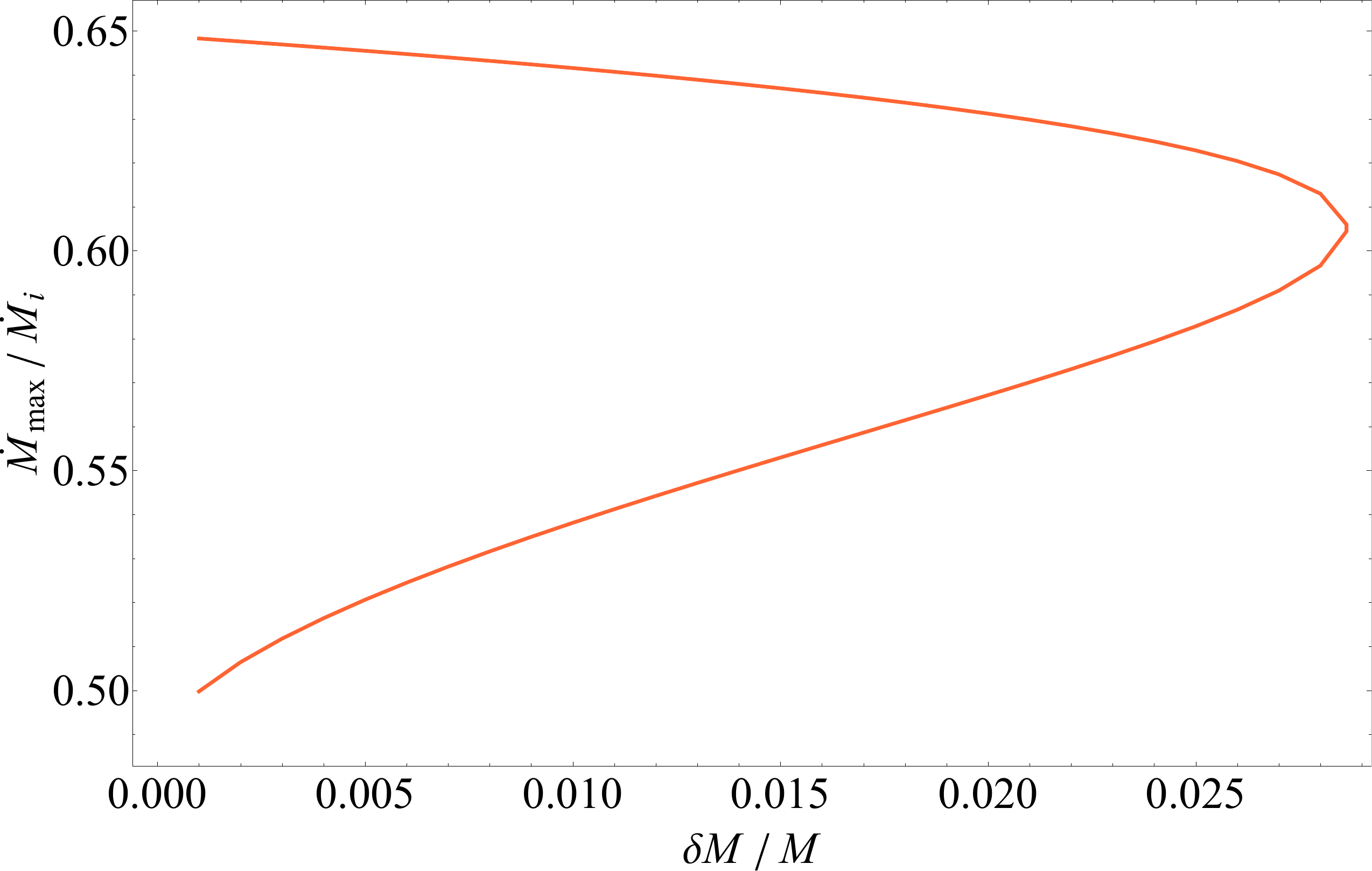} 
   \caption{The maximum accretion rate onto the black hole as a function of the fractional mass loss to neutrinos for $n = 2.5$ and $\gamma = 1.4$. There are two solutions per mass loss, one (the smaller) corresponding to the weak-shock solution, the other (larger) being the stronger-shock solution. }
   \label{fig:Mdotmax}
\end{figure}

Figure \ref{fig:Mdotmax} shows the maximum accretion rate normalized by $\dot{M}_{\rm i}$ for $n = 2.5$ and $\gamma = 1.4$ as a function of the fractional mass loss $\delta M/M$. As for the kinetic energy (Figure \ref{fig:Ekmax_shock}), there are two values of the accretion rate for a given mass loss -- the larger value corresponds to the stronger-shock solution, while the smaller is appropriate to the weak-shock solution. In general, the dimensionless ratio $\dot{M}_{\rm max}/\dot{M}_{\rm i}$ is of the order unity.

For the values from the YSG simulation in \citet{fernandez18}, we find $\dot{M}_{\rm i} \simeq 10^{-5} M_{\odot}$ s$^{-1}$. If the gas has sufficient angular momentum that it can circularize about the black hole, which is possible owing to the convective motions in the envelope of the star \citep{quataert19, antoni22}, then the energy released due to accretion (with a radiative efficiency of $0.1$) is $L_{\rm acc} \simeq 0.1\dot{M}_{\rm i} c^2 \simeq 10^{48}$ erg s$^{-1}$, which is hypercritical (i.e., many orders of magnitude above Eddington). Moreover, if the angular velocity profile (i.e., the azimuthal or poloidal velocity as a function of radius) of the envelope is decomposed into spherical harmonics, then the effect of the angular momentum on the post-shock flow can be modeled through a perturbative approach. In this way, one can account for the the angular momentum of the infalling debris self-consistently and determine the time at which a rotationally supported disc is expected to form. We leave such an analysis, and the corresponding implications for producing highly energetic astrophysical transients, to future work.

\section{Summary and Conclusions}
\label{sec:summary}
The implosion of the core of a massive star at the end of its life is accompanied by the formation of a neutron star and the liberation of $\lesssim 0.5 M_{\odot}$ of mass-energy in the form of neutrinos. Here we used a dynamical model (Section \ref{sec:basic}), in which the pressure of the gas is ignored, to show that the response of the envelope of the star (which is still causally unaware of the implosion) to this mass loss results in the formation of a caustic where fluid elements start to cross, implying that a shock will form at some radius within the star. We then showed that when the stellar envelope has a power-law density profile, there is a self-similar solution for the response of the envelope to the mass loss (Section \ref{sec:ambient}), and this self-similar solution terminates in a sonic point that expands as $\propto t^{2/3}$ -- again implying that a shock must join the outer, expanding envelope onto the inner flow that falls onto the black hole. 

Provided that the fractional mass loss is less than a critical value $\left(\delta M/M\right)_{\rm max}$ (provided in Table \ref{tab:Machmax} for different ambient power-law and adiabatic indices) and the power-law index of the ambient medium satisfies $2 < n < 3.5$, we showed in Section \ref{sec:shock} that this shock and the post-shock flow behave self-similarly, with the shock expanding as $R_{\rm sh} \propto t^{2/3}$. For a given $\delta M/M$ (below the maximum-possible value), there are two solutions that enable the smooth passage of the fluid variables through a sonic point in the interior of the flow and accretion onto the black hole. These solutions are characterized by their constant Mach number, $\mathscr{M}_{\rm sh, s}$ and $\mathscr{M}_{\rm sh, w}$, with $\mathscr{M}_{\rm sh, s} > \mathscr{M}_{\rm sh, w}$.

As described in Section \ref{sec:stability}, the stronger-shock solution with $\mathscr{M}_{\rm sh, s}$ must be unstable, while that with $\mathscr{M}_{\rm sh, w}$ is stable. Therefore, if a shock has a Mach number greater than $\mathscr{M}_{\rm sh, s}$, it will continue to strengthen into the energy-conserving/Sedov-Taylor regime, while any shock with a Mach number less than $\mathscr{M}_{\rm sh, s}$ but greater than $\mathscr{M}_{\rm sh, w}$ will weaken and asymptotically approach $\mathscr{M}_{\rm sh, w}$ (and, similarly, if a shock has a mach number less than $\mathscr{M}_{\rm sh, w}$, it will strengthen to approach $\mathscr{M}_{\rm sh, w}$). The rate at which the shock deviates from or approaches one or the other of these solutions can be derived with a perturbation analysis, which we defer to future work, but in general we expect this rate to be very slow, i.e., the initial shock position and velocity (neither of which is arbitrary for the self-similar solutions) will likely have a prolonged and pronounced effect on the shock propagation. 

In a realistic core-collapse progenitor, the density profile will exhibit variations with radius that are not captured by a single power-law, but can be approximated as a sequence of power-laws across different shell burning regions and, ultimately, the hydrogen envelope (see Figure \ref{fig:mesa_star}). For regions that have a density profile shallower than $\propto r^{-2}$, we expect the shock to asymptotically approach the weak-shock solution, while the shock Mach number will grow if $n > 3.5$ (Section \ref{sec:nle2}). In general, we expect the shock to go through phases of strengthening and weakening depending on the instantaneous Mach number and the mass loss, and the strength of the shock upon reaching the surface can likely be constrained by stitching together the strong- and weak-shock solutions in a piecewise-continuous way (Section \ref{sec:implications}). For both the weak- and strong-shock solutions, all of the gas that is hit by the shock remains bound to the black hole, implying that fallback accretion -- provided the stellar envelope has enough angular momentum to support the formation of a disc -- could produce highly energetic transients (Section \ref{sec:fallback}). 

The formation of a black hole in a failed supernova implies that the solution near the compact object should be accreting, and hence the self-similar solutions with infall described here are the most physically relevant. However, there are also ``settling'' solutions that exist between (in terms of the shock Mach number) the weak-shock and strong-shock, self-similar, accreting solutions, for which the shock velocity appoaches zero as the origin is approached. Two such settling solutions (with specific shock Mach numbers) maintain zero mass flux near the origin, and these are presented and described in Appendix \ref{sec:settling}. These solutions could be appropriate to non-terminal explosions and eruptions of massive stars, or to the response of a circumbinary disc to the mass lost to gravitational waves upon the merger of a black hole binary. The application to these other systems, as well as a comparison to numerical hydrodynamics calculations and an investigation of the strengthening and weakening of the shock, will be the subject of future work.
\newline
\newline
I acknowledge support from the National Science Foundation through grant AST-2006684 and the Oakridge Associated Universities through a Ralph E.~Powe junior faculty enhancement award.

\bibliographystyle{aasjournal}

\appendix
\section{Shock Jump conditions and Self-similar boundary conditions}
\label{sec:bcs}
The shock that joins onto the expanding envelope is expected to be weak, with a Mach number $\mathscr{M} \sim few$, and hence the pre-shock velocity (which is non-zero owing to the mass loss) and sound speed are not ignorable in comparison to the shock speed. For non-zero envelope velocity and pressure, the conservation of the mass, energy, and momentum across the shock in the comoving frame of the shock yield
\begin{equation}
\rho_{\rm s}\left(v_{\rm s}-V_{\rm sh}\right) = \rho_{\rm e}\left(v_{\rm e}-V_{\rm sh}\right),
\end{equation}
\begin{equation}
\rho_{\rm s}\left(v_{\rm s}-V_{\rm sh}\right)^2+p_{\rm s} = \rho_{\rm e}\left(v_{\rm e}-V_{\rm sh}\right)^2+p_{\rm e},
\end{equation}
\begin{equation}
\frac{1}{2}\left(v_{\rm s}-V_{\rm sh}\right)^2+\frac{\gamma}{\gamma-1}\frac{p_{\rm s}}{\rho_{\rm s}} = \frac{1}{2}\left(v_{\rm e}-V_{\rm sh}\right)^2+\frac{\gamma}{\gamma-1}\frac{p_{\rm e}}{\rho_{\rm e}}.
\end{equation}
Here subscript-e quantities refer to those of the expanding envelope, a suscript-s refers to the post-shock fluid, and $V_{\rm sh}$ is the shock speed. Solving the above three equations for the post-shock velocity, pressure, and density yields

\begin{equation}
v_{\rm s}\left(R_{\rm sh}\right) = \frac{2}{\gamma+1}\left(1+\frac{\gamma-1}{2}\frac{v_{\rm e}}{V_{\rm sh}}-\frac{\gamma p_{\rm e}}{\rho_{\rm e}V_{\rm sh}^2\left(1-\frac{v_{\rm e}}{V_{\rm sh}}\right)}\right)V_{\rm sh}
\end{equation}
\begin{equation}
p_{\rm s}(R_{\rm sh}) = \frac{2}{\gamma+1}\left(\left(1-\frac{v_{\rm e}}{V_{\rm sh}}\right)^2-\frac{\gamma-1}{2}\frac{p_{\rm e}}{\rho_{\rm e}V_{\rm sh}^2}\right)\rho_{\rm e}V_{\rm sh}^2
\end{equation}
\begin{equation}
\rho_{\rm s}(R_{\rm sh}) = \frac{\gamma+1}{\gamma-1}\left(1+\frac{2\gamma}{\gamma-1}\frac{p_{\rm e}}{\rho_{\rm e}V_{\rm sh}^2\left(1-\frac{v_{\rm e}}{V_{\rm sh}}\right)^2}\right)^{-1}\rho_{\rm e}. \label{rhobc}
\end{equation}  
The right-hand sides of each one of these expressions is evaluated at the location of the shock. When the ambient velocity ($v_{\rm e}$) and ambient pressure ($p_{\rm e}$) are much less than the shock speed and the ambient ram pressure, respectively, these manifestly reduce to the strong-shock expressions. 

Using the self-similar solutions for the envelope fluid variables (e.g., $v_{\rm e} = \sqrt{GM/r}f_{\rm e}$) and the shocked-fluid variables (e.g., $v_{\rm s} = V_{\rm sh}f_{\rm s}$), these yield the following boundary conditions for the self-similar functions $f_{\rm s}$, $g_{\rm s}$, and $h_{\rm s}$:
\begin{equation}
f_{\rm s}(1) = \frac{2}{\gamma+1}+\frac{\gamma-1}{\gamma+1}\frac{3\eta_{\rm sh}}{2}f_{\rm e}(\eta_{\rm sh}) 
-\frac{2\gamma}{\left(\gamma+1\right)\left(n+1\right)}\frac{9\eta_{\rm sh}^2}{4}\frac{h_{\rm e}(\eta_{\rm sh})}{g_{\rm e}(\eta_{\rm sh})}\left(1-\frac{3\eta_{\rm sh}}{2}f_{\rm e}(\eta_{\rm sh})\right)^{-1}, \label{bcfs}
\end{equation}
\begin{equation}
h_s(1) = \bigg(\frac{2}{\gamma+1}\left(1-\frac{3\eta_{\rm sh}}{2}f_{\rm e}(\eta_{\rm sh})\right)^2 
-\frac{\gamma-1}{\gamma+1}\frac{9\eta_{\rm sh}^2}{4}\frac{1}{n+1}\frac{h_{\rm e}(\eta_{\rm sh})}{g_{\rm e}(\eta_{\rm sh})}\bigg)g_{\rm e}(\eta_{\rm sh}), \label{bchs}
\end{equation}
\begin{equation}
g_{\rm s}(1) = \frac{\gamma+1}{\gamma-1}\bigg(1+\frac{2\gamma}{\gamma-1}\frac{9\eta_{\rm sh}^2}{4}\frac{1}{n+1}\frac{h_{\rm e}(\eta_{\rm sh})}{g_{\rm e}(\eta_{\rm sh})} \left(1-\frac{3\eta_{\rm sh}}{2}f_{\rm e}(\eta_{\rm sh})\right)^{-2}\bigg)^{-1}g_{\rm e}(\eta_{\rm sh}). \label{bcgs}
\end{equation}
Upon specifying the value of $\eta_{\rm sh}$, these boundary conditions can be used to integrate the self-similar equations (Equations \ref{ss1} -- \ref{ss3}) from $\xi = 1$ inward, and the values of $\eta_{\rm sh}$ are those that maintain the continuity of the fluid variables through a sonic point in the flow.

\section{Settling solutions}
\label{sec:settling}
The self-similar solutions described in Section \ref{sec:shock} match onto the boundary conditions \eqref{bcfs} -- \eqref{bcgs} and yield accretion onto the newly formed black hole at the origin. There are two such solutions with dimensionless shock parameters $\eta_{\rm sh, s}$ and $\eta_{\rm sh, w}$, such that $\eta_{\rm sh, s} < \eta_{\rm sh, w}$ (i.e., the strong-shock velocity is larger than the weak-shock velocity). As described in Section \ref{sec:discussion}, these values of $\eta_{\rm sh}$ are constrained by requiring that the fluid variables smoothly pass through a sonic point in the interior of the flow, which are also known as type-II similarity solutions \citep{sedov59}. The sonic condition is 
\begin{equation}
\left(f_{\rm s}-\xi\right)^2g_{\rm s}-\gamma h_{\rm s} = 0. \label{sonic}
\end{equation}

There are also solutions to the fluid equations with $\eta_{\rm sh, s} < \eta_{\rm sh} < \eta_{\rm sh, w}$ that \emph{settle} onto the origin and are causally connected everywhere, meaning that $\left(f_{\rm s}-\xi\right)^2g_{\rm s}-\gamma h_{\rm s} < 0$ for all $\xi$. Specifically, all solutions with $\eta_{\rm sh, s} < \eta_{\rm sh} < \eta_{\rm sh, w}$ have $f_{\rm s} \rightarrow 0$ in the limit that $\xi \rightarrow 0$, and there are two specific values of $\eta_{\rm sh}$ that have zero mass flux near the origin, i.e., $f_{\rm s}g_{\rm s}\xi^2\rightarrow 0$ as $\xi \rightarrow 0$. We denote the two values of $\eta_{\rm sh}$ that satisfy this additional condition on the mass flux by $\tilde{\eta}_{\rm sh, s}$ and $\tilde{\eta}_{\rm sh, w}$, with $\tilde{\eta}_{\rm sh, s} < \tilde{\eta}_{\rm sh, w}$, and similarly for the self-similar functions (i.e., $\tilde{f}_{\rm s, s}$ and $\tilde{f}_{\rm s, w}$ are the strong and weak settling solutions, respectively, and analogously for $g_{\rm s}$ and $h_{\rm s}$). 

\begin{figure}[htbp] 
   \centering
   \includegraphics[width=0.495\textwidth]{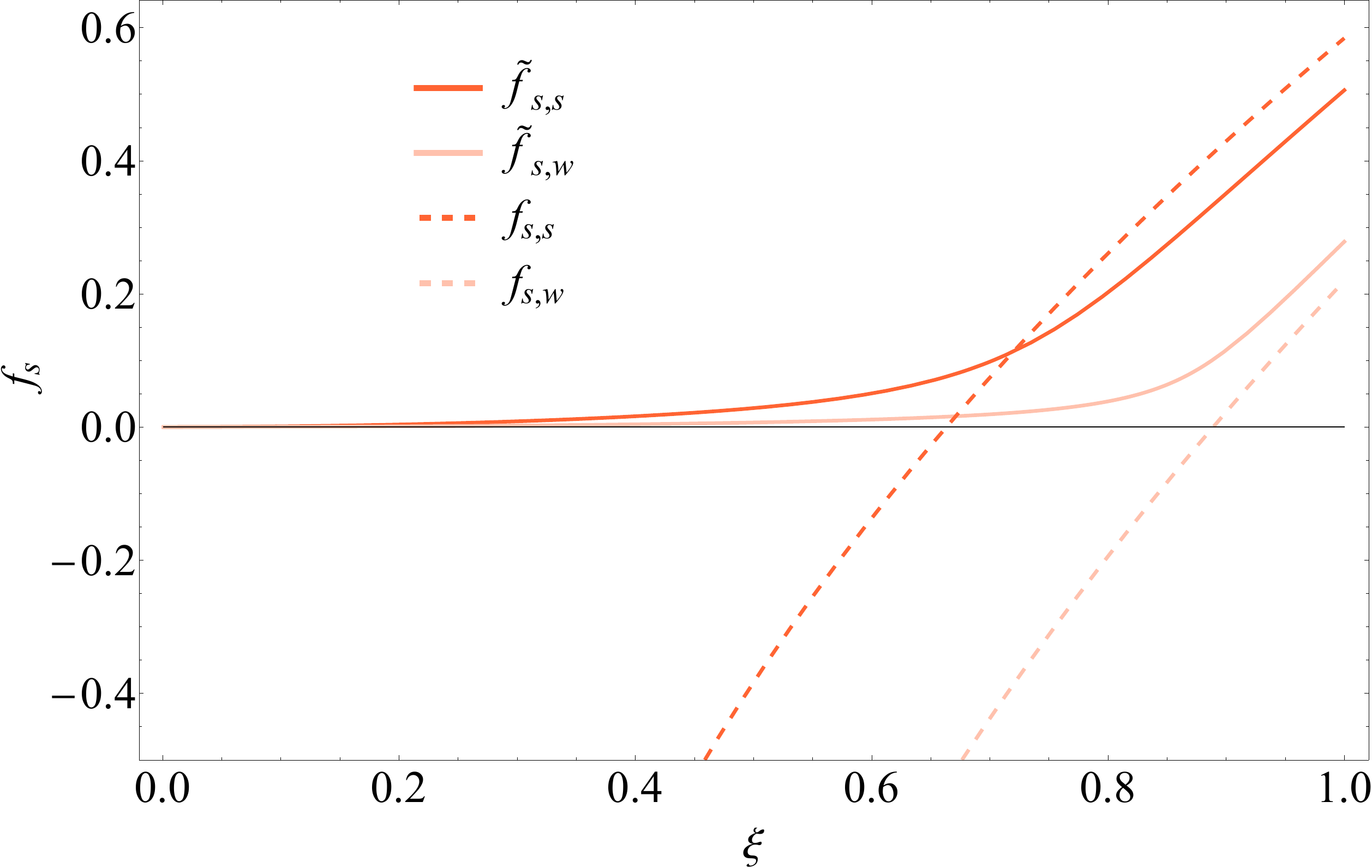} 
  \includegraphics[width=0.495\textwidth]{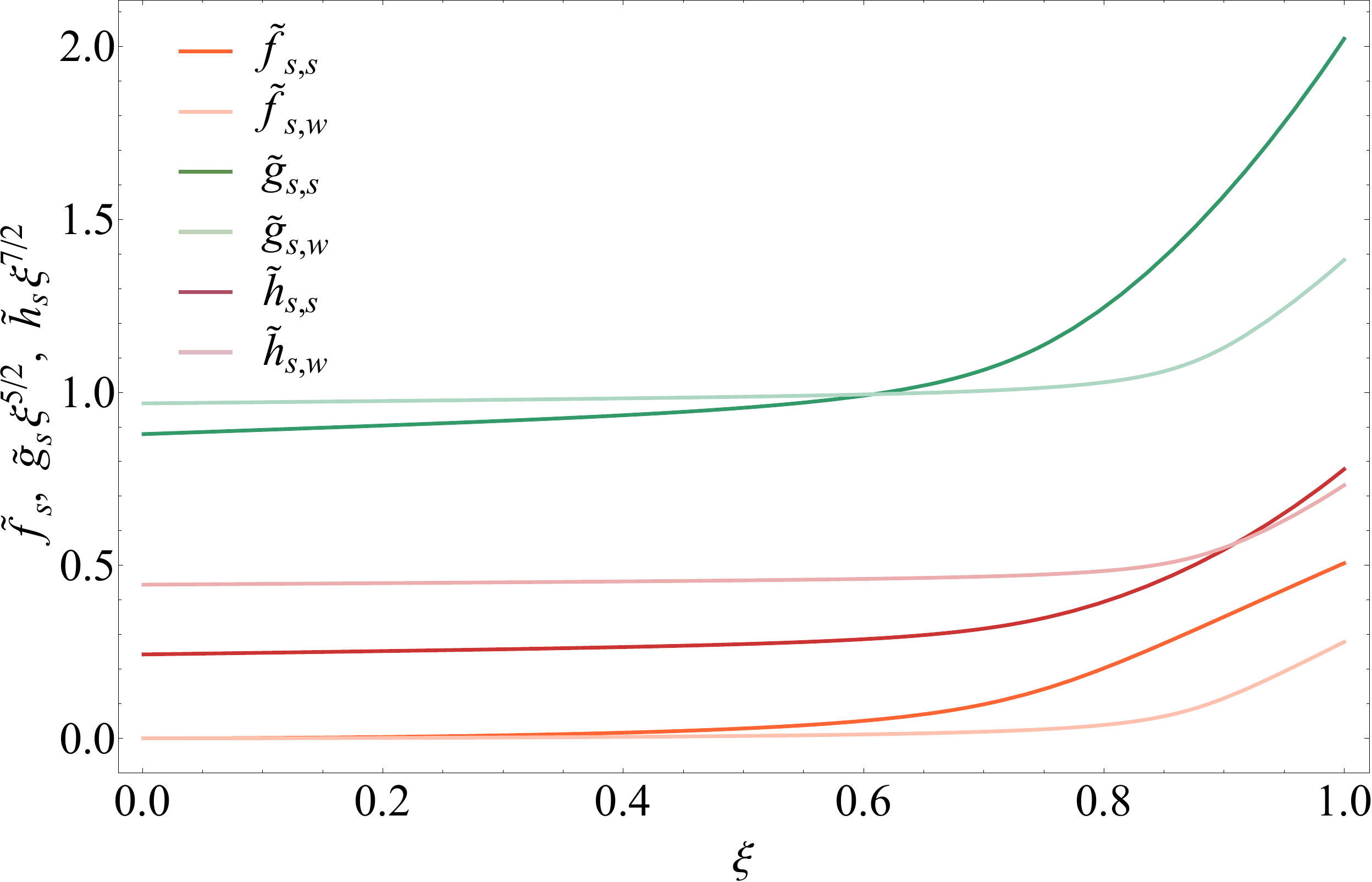} 
   \caption{Left: The settling (solid lines) and accreting (dashed lines) post-shock solutions for $n = 2.5$, $\gamma = 1.4$, and $\delta M/M = 0.01$. The dark (light) curves are the strong-shock (weak-shock) solutions. Right: The strong-shock (dark) and weak-shock (light) settling solutions for the self-similar velocity, density, and pressure for $n = 2.5$, $\gamma = 1.4$, and $\delta M/M = 0.01$. The dimensionless density varies as $\tilde{g}_{\rm s} \propto \xi^{-2.5}$ near the origin, while the pressure varies as $\tilde{h}_{\rm s} \propto \xi^{-3.5}$. }
   \label{fig:settle}
\end{figure}

The left panel of Figure \ref{fig:settle} shows the accreting solutions (dashed) and the settling solutions (solid) for the fiducial case of $n = 2.5$, $\gamma = 1.4$, and $\delta M/M = 0.01$, for which we find $\tilde{\eta}_{\rm sh, s} \simeq 0.658$ and $\tilde{\eta}_{\rm sh, w} \simeq 0.849$. The strong (weak) solution is shown by the dark (light) curve. Unlike the accreting solutions, which have zero velocity at a given $\xi$ and within that radius approach freefall onto the black hole at the origin, the settling velocity remains positive everywhere and approaches zero as $\tilde{f}_{\rm s} \propto \xi^{2.5}$ in the limit that $\xi \rightarrow 0$. The right panel shows the self-similar velocity, density, and pressure for the same parameters as the left panel for both the strong (dark) and weak (light) solutions. In the limit that $\xi \rightarrow 0$, we find that the self-similar density scales as $\tilde{g}_{\rm s} \propto \xi^{-2.5}$ and the pressure as $\tilde{h}_{\rm s} \propto \xi^{-3.5}$, which implies that the density and pressure approach the time-independent power-laws $\rho \propto r^{-2.5}$ and $p \propto \propto r^{-3.5}$ as $r\rightarrow 0$. At small radii, these solutions therefore conform to hydrostatic envelopes that surround the compact object.

\begin{figure}[htbp] 
   \centering
   \includegraphics[width=0.995\textwidth]{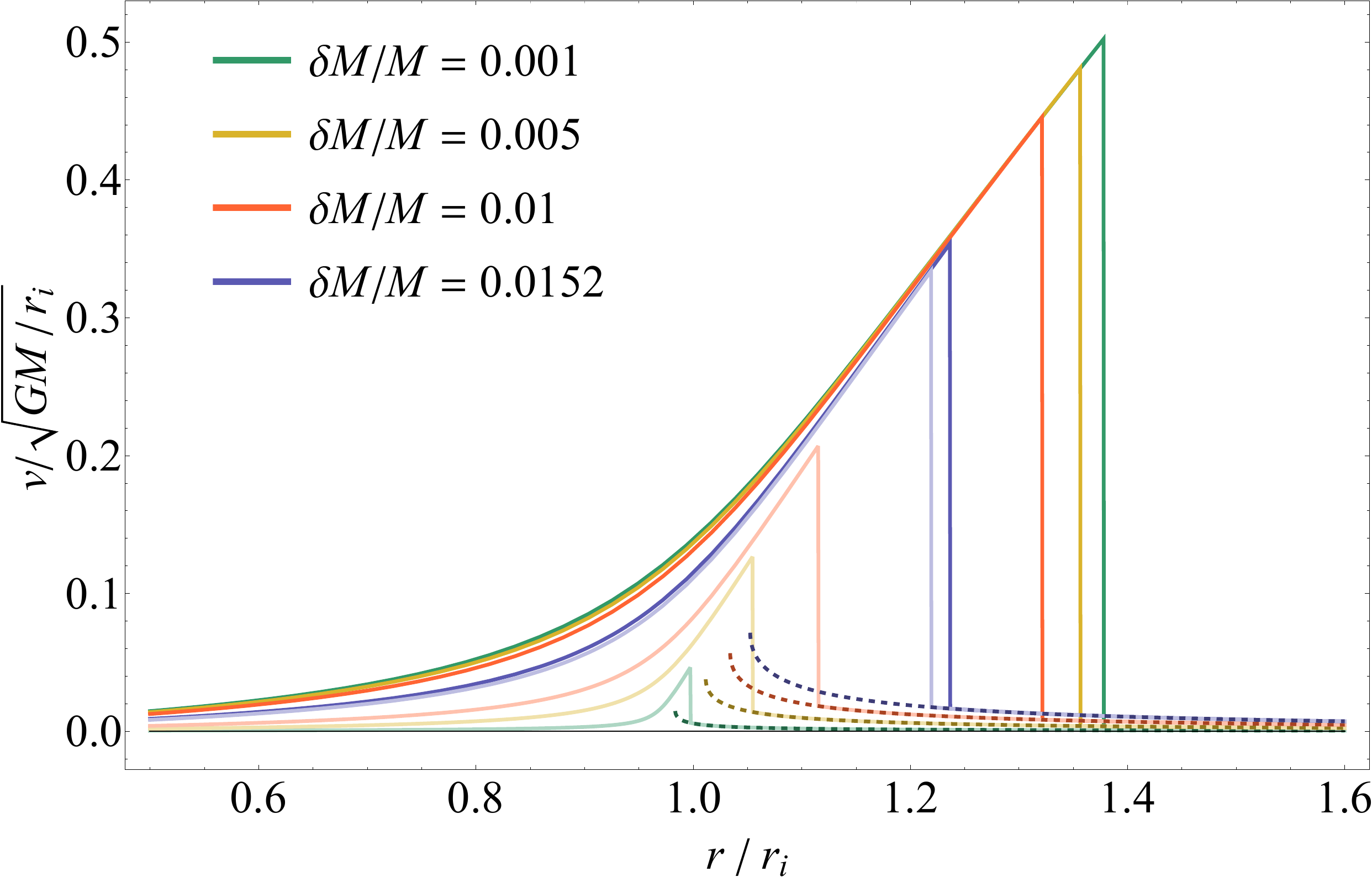} 
   \caption{The fluid velocity normalized by $\sqrt{GM/r_{\rm i}}$ as a function of radius normalized by $r_{\rm i}$ for $n = 2.5$, $\gamma = 1.4$, and the fractional mass losses in the legend. The dark curves show the strong-shock solution, the light curves show the weak-shock solution, and the dashed curves show the expanding ambient medium/envelope (onto which the shock joins). As $\delta M/M \rightarrow 0$, the weak-shock solution approaches the hydrostatic limit, i.e., for which there is no shock. The weak- and strong-shock solutions approach one another as $\delta M/M$ increases, and above $\delta M/M \simeq 0.01523$ there are no settling solutions that have zero mass flux at the origin.}
   \label{fig:v_of_r_settle}
\end{figure}

As for the accreting solutions, there are two solutions while the mass loss is below a critical value, and as $\delta M/M$ increases the two solutions approach one another and do not exist above a critical mass loss. Figure \ref{fig:v_of_r_settle} shows the velocity (normalized by $\sqrt{GM/r_{\rm i}}$) as a function of radius (normalized by $r_{\rm i}$) for the strong (dark) and weak (light) settling solutions, as well as the solution for the expanding envelope (dashed), for $n = 2.5$, $\gamma = 1.4$, and the fractional mass losses contained in the legend. As $\delta M/M$ approaches zero, the weak solution approaches the hydrostatic (i.e., no-shock) limit, and as $\delta M/M$ increases the weak-shock solution becomes stronger (in terms of the Mach number) and the strong-shock solution becomes weaker. As $\delta M/M$ approaches a limiting value of $\simeq 0.01523$, the weak and strong-shock solutions converge toward one another, and above this mass loss there are no solutions that settle onto the origin with zero mass flux. 

The interpretation of these solutions in the context of stability is analogous to that of the accreting solutions: because there are two solutions for a given set of parameters (provided $2<n<3.5$), the smaller of these is the dynamically stable one, while the larger is unstable. In terms of physical applicability, these solutions should trace the physical evolution of the fluid in weak explosions/eruptions in which there is a mechanism responsible for supplying the pressure gradient near the origin that allows the fluid to settle. One such scenario would be a non-terminal stellar eruption involving the deposition of energy through, e.g., convective wave heating \citep{shiode14}. In particular, if a wave is excited at the base of the envelope of a massive star that then steepens into a shock, we would expect the shock to continue to strengthen into the strong regime if the Mach number is larger than the one appropriate to the unstable settling solution with $\delta M/M = 0$; for $n = 2.5$ and $\gamma = 1.4$, this Mach number is $\simeq 1.71$. If the Mach number is weaker than this, then we would expect the shock to asymptotically decline in strength, leaving the gas in a hydrostatic state at sufficiently late times. 

A second system to which these solutions may apply is the gas surrounding a black hole binary at the time of merger. In particular, as two black holes merge they can liberate up to $\sim 10\%$ of the rest mass of the binary in the form of gravitational waves (e.g., \citealt{mroue13}). Previous investigations (e.g., \citealt{rossi10, corrales10, rosotti12, demink17, martin18}) have focused largely on the scenario in which the binary is surrounded by a cold, geometrically thin disc of gas, which then responds dynamically to the mass lost to gravitational waves and the kick imparted to the black hole remnant. The kick generates shocks within the disc where particle orbits cross, the energy from which is likely dissipated in highly optically-thick regions near the midplane that then radiate from the disc surface at the Eddington limit at most. 

However, it may also be possible that the gas surrounding the binary at the time of merger is more geometrically thick and pressurized, particularly if the gas is in the form of an advection dominated flow or adiabatic inflow-outflow disc \citep{narayan94, blandford99}, or if mass from the circumbinary disc is able to supply mass to the black holes until the time of merger (e.g., \citealt{artymowicz94, miranda17, duffell20, heath20, li22, west22}. If this is the case, then the response of the surrounding gas could be modeled with the settling solutions here, the assumption being that rotational support -- while (by assumption) ignorable at large radii -- provides the effective agent that halts the inflow at small radii.

\end{document}